\documentclass[runningheads]{llncs}

\usepackage{eccv}

\usepackage{eccvabbrv}

\usepackage{graphicx}
\usepackage{booktabs}
\usepackage{wrapfig}
\usepackage{graphicx}
\usepackage{dblfloatfix}
\usepackage{pifont} 
\usepackage{microtype}

\usepackage[accsupp]{axessibility}  % Improves PDF readability for those with disabilities.

\usepackage{hyperref}

\usepackage{orcidlink}

\newcommand{\taskname}{VisAH}
\newcommand{\modelname}{VisAH-FM}

\title{Conditional Flow Matching for Visually-Guided Acoustic Highlighting}

\author{
Hugo Malard\inst{1}\thanks{Work done during an internship at Meta.}
\and
Gael Le Lan\inst{2}
\and
Daniel Wong\inst{2}
\and
David Lou Alon\inst{2}
\and
Yi-Chiao Wu\inst{2}
\and
Sanjeel Parekh\inst{2}
}

\authorrunning{H. Malard et al.}

\institute{
LTCI, Télécom Paris, Institut Polytechnique de Paris
\and
Meta
}

\begin{document}
\maketitle

\begin{abstract}

Visually-guided acoustic highlighting seeks to rebalance audio in alignment with the accompanying video, creating a coherent audio–visual experience. While visual saliency and enhancement have been widely studied, acoustic highlighting remains underexplored, often leading to misalignment between visual and auditory focus. Existing approaches use discriminative models, which struggle with the inherent ambiguity in audio remixing, where no natural one-to-one mapping exists between poorly-balanced and well-balanced audio mixes.
To address this limitation, we reframe this task as a generative problem and introduce a Conditional Flow Matching (CFM) framework. A key challenge in  iterative flow-based generation is that early prediction errors — in selecting the correct source to enhance —
compound over steps and push trajectories off-manifold. To address this, we introduce a rollout loss that penalizes drift at the final step, encouraging self-correcting trajectories and stabilizing long-range flow integration. We further propose a conditioning module that fuses audio and visual cues before vector field regression, enabling explicit cross-modal source selection.
Extensive quantitative and qualitative evaluations show that our method consistently surpasses the previous state-of-the-art discriminative approach, establishing that visually-guided audio remixing is best addressed through generative modeling. Qualitative samples are available at the \href{https://hugomalard.github.io/visah-fm/#}{project page.}
\end{abstract}    
\section{Introduction}
\label{sec:intro}
%\paragraph{Why is smart remixing important}
% \SP{Our story:
% - We solve VisAH; why is it important
% - Inherent property of the task and recent results from generative modeling point towards a generative formulation -> flow matching
% - Important to understand what to enhance. Two ways of attacking it:
% — rollout loss to generate more accurate tranjectories
% — conditioning module to provide more explicit cues for drift estimation}
The recent proliferation of video content creation and consumption underscores the need for meticulous curation of both visual and audio elements to deliver an engaging user experience. While visual cue manipulation—through techniques such as optimal viewpoint selection or post-editing—has been a long-standing focus in media production \cite{lei21,lin23,liu22}, the audio domain has not seen equivalent advancements. This disparity frequently results in a noticeable disconnect between visual saliency and acoustic emphasis.
\begin{figure}[!t]
        \centering
        \includegraphics[width=0.95\linewidth]{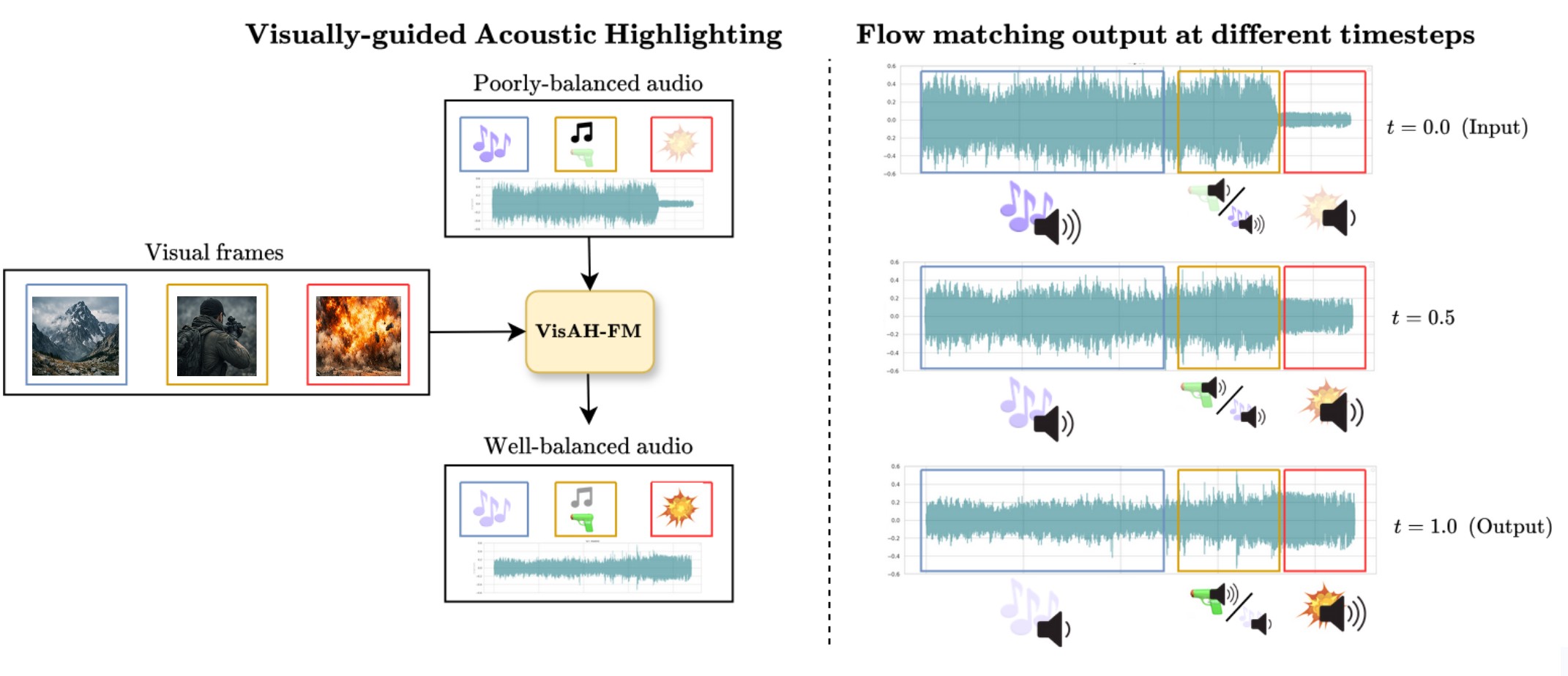}
        \caption{VisAH-FM casts visually-guided acoustic highlighting as flow matching: sources are iteratively enhanced when audio-visual cues agree and subdued otherwise.}
    \label{fig:archi}
\end{figure}
For instance, when a person speaks on camera, viewers intuitively expect their voice to be clear and prominent. However, without deliberate mixing, their speech may be overwhelmed by background noise or music, breaking the perceptual alignment between what is seen and what is heard.
Raw daily recordings, in particular, often contain poorly balanced audio due to recording device limitations where microphones attached to video cameras capture all sounds indiscriminately. 

To bridge this crucial gap, the task of Visually-guided Acoustic Highlighting (VisAH) was introduced \cite{visah}, aiming to automatically transform poorly balanced audio by using the accompanying video stream as guidance, ultimately creating a more harmonious audio-visual experience. This process involves rebalancing loudness of different audio sources, like speech, music, and sound effects, to reflect their relative prominence as implied by the visual context.
Previously, this task  was approached using a  discriminative paradigm~\cite{visah}, relying on sound source separation architectures, such as DEMUCS~\cite{Hdemucs}. Thus, training for a one-to-one mapping. However, we observe that given the video stream, there are several ways of creating poor or well-balanced audio mixes. In other words, audio remixing (or rebalancing) is fundamentally a task of transporting from one distribution to another. Given its many-to-many nature we argue that a generative approach is better suited than a purely discriminative one. Indeed, prior work~\cite{ng01} shows that in many-to-many settings, where both input and output contain mutual uncertainty, generative models can exploit information from the marginal to capture correlations and latent structure that discriminative models ignore. Additionally, recent works have shown that, for multiple audio tasks where there is no clear one-to-one mapping, like speech enhancement or audio source separation, generative approaches significantly outperform their discriminative counterparts~\cite{davis,FMsource,DAVSE,musicSep}.

%\paragraph{Why flow matching is challenging in this setting: early steps are harder and errors accumulate.}
Among generative modeling techniques, flow matching~\cite{lipman2022flow} provides a principled framework for learning a continuous transformation between input and target distributions, making it a natural fit for remixing as shown in Figure~\ref{fig:archi}. 
%Additionnaly, studies~\cite{LBM} have shown that using flow (or bridge) matching
In this formulation, a neural network learns the vector field that guides samples from the source distribution towards the target through a sequence of intermediate states.
For the visually guided acoustic highlighting task, the main difficulty lies in identifying the audio sources to enhance or suppress. Ideally, this crucial operation must occur from the very first step of the flow matching process, where the model begins transforming the input distribution towards the target one.
% \SP{This is nice, we can avoid being very focused on first step and mention here that we confirm this hypothesis through experiments and ablation studies.}
As this step is inherently difficult, errors, even if small, can easily arise early on. In an iterative framework like flow matching, such early inaccuracies propagate through subsequent steps, compounding over time and driving the trajectory away from the data manifold. This accumulation of error substantially degrades model performance and motivates our proposal of the \textit{rollout loss}, an additional loss to stabilize predictions across steps.
%\paragraph{Conditioning: the need to fuse audio and visual information before flow matching.}

Furthermore, the source identification problem heavily relies on the conditioning of the model. In prior work, this was based solely on visual cues—either features extracted from an image encoder or textual captions describing the entire visual scene. 
%Applying the same conditioning in a model repurposed for the flow matching task
%These image-only features were then passed directly to the \textit{drift estimator}—the core flow-matching network responsible for learning the transport vector field that maps input audio distributions to their remixed counterparts. As a result, the drift estimator had to implicitly learn the correspondence between auditory and visual signals, a task that goes beyond its primary objective of modeling the flow dynamics.
These image-only features were passed through a small transformer and then fed to the main denoising model. Consequently, the latter had to implicitly learn the correspondence between auditory and visual signals—a task that goes beyond its primary goal of regressing the audio from a conditioning signal.
We argue that this ``division of responsibilities" can be made more effective. By integrating audio features directly into the visual encoder, we enable cross-modal filtering within the conditioning pathway itself. Such early fusion allows the conditioning network to preserve only the visual components relevant to the target sound sources, yielding a compact, audio-aware representation. The main model then receives conditioning features that are already semantically aligned with the audio content, allowing it to ``concentrate" entirely on accurate audio regression rather than on discovering cross-modal correspondences.
%We argue that this separation of concerns can be made more effective. By integrating audio features directly into the visual encoder, we enable \textit{cross-modal filtering} within the conditioning module itself. This early audio integration allows the conditioning pathway to retain only the visual components relevant to the target sound source, producing a compact, audio-aware representation. The drift estimator then receives conditioning features that are already semantically aligned with the audio content, allowing it to focus solely on learning the vector field for distribution transport.

%This produces more explicit and task-aligned conditioning signals for the flow matching module.

%\paragraph{How we solve it and our contributions and findings}
In summary, our contributions are threefold:
\begin{itemize}
\item We reformulate the visually guided acoustic highlighting problem as a \textit{generative} task rather than a discriminative one, and demonstrate the advantages of this formulation both qualitatively and quantitatively.
\item We introduce a \textit{rollout loss} that mitigates the error accumulation inherent to standard flow matching for such a task, and analyze its impact in detail.
\item We design an improved \textit{conditioning module} that enables earlier fusion of audio and visual features before the velocity field estimation stage, improving performance and achieving state-of-the-art results.
\end{itemize}

\iffalse
\begin{figure*}[t]
    \centering
    % ---- First subfigure ----
    \begin{subfigure}[t]{0.48\linewidth}
        \centering
        \includegraphics[width=\linewidth]{mainFigA.png}
        \caption{VisAH-FM repurposes VisAH for flow matching by adding timestep conditioning in the latent space and recurrently estimating trajectories.}
        \label{fig:archi}
    \end{subfigure}
    \hfill
    % ---- Second subfigure ----
    \begin{subfigure}[t]{0.4\linewidth}
        \centering
        \includegraphics[width=\linewidth]{mainFigB.png}
        \caption{Conditioning module: CLIP and auxiliary features are projected, cross-attended, projected back, and added to the original CLIP features.}
        \label{fig:condDetailed}
    \end{subfigure}
    
    \caption{Proposed method: (a) global architecture and (b) conditioning module.}
    \label{fig:condModules}
\end{figure*}
\fi
\section{Proposed Method}
\label{sec:method}
\subsection{Task Formulation}
\paragraph{Visually-guided Acoustic Highlighting.}
The visually-guided acoustic highlighting task was introduced to address the disconnect often observed between visual and acoustic saliency in video content. Its objective is to utilize the accompanying video stream as guidance to transform a poorly balanced audio track into a well-balanced one exhibiting appropriate highlighting effects. To facilitate training models for this task, the Muddy Mix Dataset \cite{visah} was curated, leveraging the  meticulous audio-visual crafting found in movies, providing a form of "free supervision". The input audio, representing poorly mixed content is generated using a pseudo-data generation process that simulates real-world mixing deficiencies. This process systematically disturbs the original high-quality movie audio through three key steps: (1) Separation, decomposing the audio into source components (speech, music, and sound effects); (2) Adjustment, altering the relative levels of these separated sources by applying suppression or emphasis using selected strength levels (high, moderate, low); and (3) Remixing, linearly combining the adjusted sources to form the ill-balanced input audio.
%While the existing pseudo-data generation paradigm is suitable for simulating poorly mixed audio, we argue that mixing deficiencies encountered in real-world scenarios are typically characterized by continuous changes in loudness rather than strictly discrete, hard increases or decreases in source levels. Therefore, we revised the dataset generation strategy by replacing the discrete categorization with a uniform sampling method over a continuous range of decibel values (6 and 12), thereby introducing greater variability while following the fundamental procedures established by Muddy Mix.

\subsection{Our Approach}

\paragraph{Conditional Flow Matching (CFM).}
Unlike discriminative models, which learn a point-wise direct one-to-one mapping between inputs and targets, flow matching methods focus on aligning probability distributions. They model a continuous transformation between a source and target distribution, enabling many-to-many correspondences rather than explicit pairwise matches.
%\GL{Could be worth explaining what $x_0$ and $x_1$ (dimensions...) are, especially considering one of them is usually from the standard distribution.}
Given two distributions from which we can sample during training, $(x_0, x_1) \sim \pi_0 \times \pi_1$, where $x_0 \in \mathbb{R}^d$ is typically drawn from a base distribution (e.g., a standard Gaussian) and $x_1 \in \mathbb{R}^d$ represents a sample from the target data distribution, flow matching learns a time-dependent velocity field that continuously transports samples from $\pi_0$ to $\pi_1$ \cite{lipman2022flow,tong2023conditional}. 
An interpolant between the two samples is defined as
%Given two distributions from which we can sample during training, $(x_0\in \mathbb{R}^{n_t \times d_1 }, x_1 \in \mathbb{R}^{n_t \times d_2 }) \sim \pi_0 \times \pi_1$
\begin{equation}
    x_t = (1-t)x_0 + t x_1,
    \label{eq:interpolant}
\end{equation}
whose evolution follows the conditional ODE
\begin{equation}
    \frac{dx_t}{dt} = \frac{x_1 - x_t}{1 - t}.
\end{equation}
In practice, a neural network $v_\theta(x_t, t)$ is trained to approximate this velocity field field by minimizing
\begin{equation}
    \mathcal{L}_{\text{CFM}}(\theta) =
    \mathbb{E}_{t, x_0, x_1}
    \left[\left\|\frac{x_1 - x_t}{1 - t} - v_\theta(x_t, t)\right\|^2_2\right].
    \label{eq:objFM}
\end{equation}

\paragraph{Conditional Flow Matching for Acoustic Highlighting.}
We propose to cast the acoustic highlighting problem as an instance of conditional flow matching, where the model has to learn the flow that goes from a poorly-balanced to a well-balanced audio distribution, while being conditioned on the visual cues. 
We argue that this formulation is particularly well-suited to the task, as a given video may admit multiple plausible high-quality mixes, as well as diverse degraded versions. Consequently, matching distributions rather than individual instances better captures the underlying variability.
%We argue that this setup suits more the task, as there is more than a single good remixing for a video, and that there could be multiple degradation (poorly mix) possible for a given video. Hence the distribution matching approach of flow matching seems more suited. In fact, we sample on-the-fly the sources to enhance and diminish, making the model see multiple different inputs for the same output across epochs.
More formally, plugging into the conditional flow matching setup mentioned above, we are trying to minimize:
\begin{equation}
    \mathcal{L}_{\text{CFM}}(\theta) =\mathbb{E}_{t,x_0,x_1}[\|(x_1-x_t)/(1-t)-v_{\theta}(x_t,t,c)\|^2_2]
\end{equation}
Here $x_0$ and $x_1$ are audios from poorly-balanced and well-balanced audio distributions, respectively, $c$ is the visual conditioning, and $t\sim \mathcal{U}(0,1)$ the sampled timestep. During training, we expose the model to multiple distinct inputs corresponding to the same target across epochs. We do so by randomly generating  poorly balanced audio samples on-the-fly as opposed to a fixed dataset used in \cite{visah}.

Also, motivated by prior work that shows flow matching benefits from pretrained discriminative initialization~\cite{flowdec,storm}, we adapt the VisAH architecture \cite{visah} and training objective to the generative setting while initializing with its pretrained weights. Additional implementation details in Sec.~\ref{sec:exp}.% \GL{maybe some maths to explain that? Or just mention 3.1}.

\subsubsection{CFM Training with Rollout Loss.}%\GL{Use the term backpropagation through the flow/solver?}

% \subsubsection{Training objectives}
%In a regular flow matching setup, during training, each step is assumed to be independent from the others, and the ground truth input is fed to the model (i.e., the real $x_1$ is used to obtain the input $x_t$). However, in the acoustic highlighting task, the core part consists of finding which source should be enhanced/decreased. This crucial part should be performed in the very first steps, resulting in difficulty discrepancies between steps (i.e., last steps are easier).

In a standard flow matching setup, each step is trained independently, and the ground-truth sample is used as input at every time step (i.e., the real $x_0$ is used to obtain $x_t$).
%\GL{For the next paragraph I would suggest remaining generic by just saying that flow matching can inherently generate errors that compound throughout the trajectory. The term drift would be more appropriate in that context rather than the one you defined earlier. To overcome that drift you propose to add backpropagation through the whole trajectory: not only teaching the model what velocity it should predict at any given point but also making sure it self corrects to reach the target point.}

However, flow matching can inherently generate drift along the predicted trajectory: small local inaccuracies in the velocity field may compound over successive steps, gradually deviating the model’s trajectory from the target distribution. To mitigate this drift, we propose using backpropagation through the flow, enabling the model not only to learn the correct instantaneous velocity at each step but also to self-correct so that the overall trajectory converges toward the target endpoint.

In the acoustic highlighting task, this effect is particularly pronounced. The key challenge lies in determining which sources should be enhanced or attenuated—a decision that must be made during the earliest steps. Later steps primarily refine these earlier decisions. This asymmetry in step difficulty means that early-step prediction errors can have a disproportionate effect, propagating and amplifying throughout the entire trajectory.

To make the model more robust to such drift, we introduce a rollout loss that supervises the entire generated trajectory. Unlike typical flow matching applications that use hundreds of integration steps, acoustic highlighting operates on relatively close input–output distributions, allowing us to use only a few steps (four in our experiments, following \cite{LBM}). This setup makes end-to-end backpropagation through the full flow computationally feasible, enabling the model to jointly optimize intermediate updates and final consistency.

\begin{figure}
    \centering
    \includegraphics[scale=0.25]{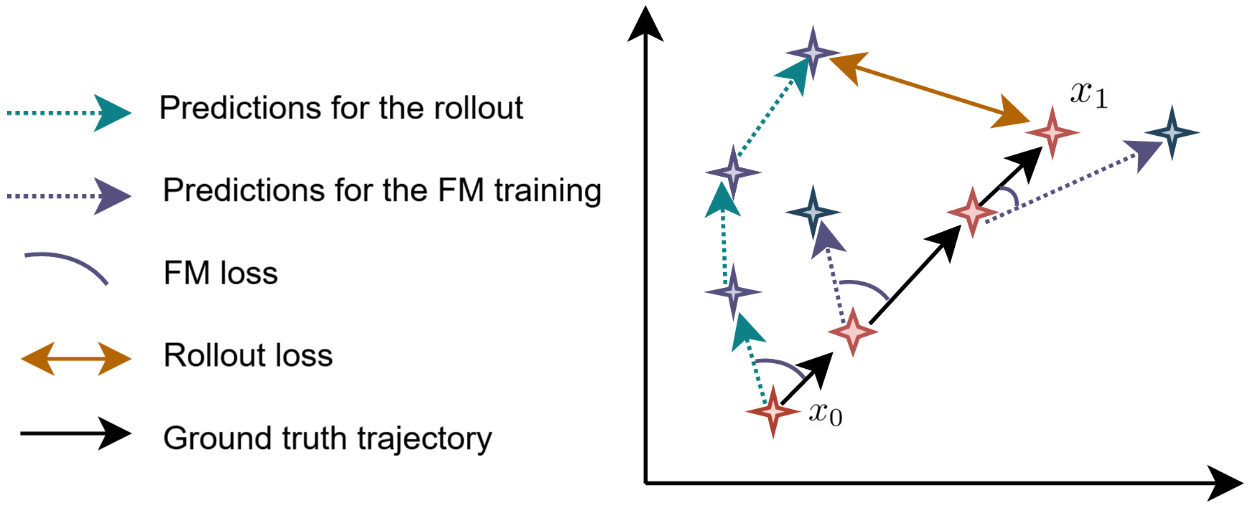}
    \caption{Rollout loss supervises the output after all flow-matching steps, encouraging coherent trajectories and reducing compounding errors.}
    \label{fig:rolloutScheme}
\end{figure}

Specifically, we perform a full rollout during training and add an auxiliary mean-squared-error (MSE) loss between the final predicted output and the ground-truth target (see Figure~\ref{fig:rolloutScheme}). This encourages consistency across steps and stability under self-generated predictions, effectively reducing the impact of early-step errors. The overall training objective is:
\begin{equation}
    \mathcal{L}(\theta)=\mathcal{L}_{\text{CFM}}(\theta) + \lambda \mathbb{E}_{x_0}[||\hat{x}_T-x_T||^2_2].%||v_{\theta}(v_{\theta}(...v_{\theta}(x_0,0),x_{n-1},n-1),x_n,n)-x_1||
    \label{eq:finalLoss}
\end{equation}
Where $\hat{x}_T$ denotes the model prediction after $T$ recurrent applications of the flow operator starting from $x_0$, and $x_T$ is the corresponding ground-truth target. Unless otherwise stated, we set $T=4$ steps (matching the total number of inference steps). 
%\vspace{-1em}
\paragraph{Relation to Consistency Models.}
Our rollout loss conceptually resembles strategies proposed in consistency models \cite{songConsistency,song2023improved} for diffusion or flow-based generation, which aim to stabilize iterative predictions by enforcing agreement between nearby timesteps.
%\GL{You can also mention "D-Flow: Differentiating through Flows for Controlled Generation" and say that you optimize the model weights rather than the model input.}
However, unlike consistency models that match predictions across infinitesimal noise scales or time intervals, our rollout loss supervises the final prediction after multiple recurrent steps against the ground truth trajectory.
This makes it focus more on the exposure-bias mitigation~\cite{exposureDif,exposureOri}: the network is explicitly trained to recover from its own intermediate errors rather than only ensuring local consistency.
The optimization of this objective is performed via backpropagation through the flow, akin in mechanism to D-Flow \cite{DFlow}.
Unlike the latter—which uses this process to optimize the model input—we use it to optimize the model weights.
Empirically, we found that this end-to-end constraint yields more stable long-range trajectories and avoids the error-amplification issues observed when applying standard consistency loss (see Table~\ref{tab:comparison_rollout}).
\subsubsection{Conditioning Module.}

%----original text------
%The VisAH model \cite{visah} builds upon a DEMUCS~\cite{demucs} architecture, conditioned in its latent space via a transformer-based conditioning module. Frame-wise features extracted from a CLIP vision encoder~\cite{CLIP} or from caption embeddings (T5~\cite{T5}) are processed by a temporal transformer and cross-attended into the U-Net latent space.

%------revised--------
As shown in Figure~\ref{fig:condDetailed}, we build upon the VisAH backbone~\cite{visah}, which consists of dual U-Net branches processing spectrogram and waveform representations, inspired by HybridDemucs~\cite{Hdemucs}. Conditioning is performed in the latent space through a transformer-based module. In VisAH, frame-wise features extracted from a CLIP vision encoder~\cite{CLIP} or from caption embeddings (T5~\cite{T5}) are processed by a temporal transformer and cross-attended into the U-Net latent space.
%However, the source to be enhanced usually corresponds to the main common concept between the audio and the image. As the conditioning module only accesses the visual information, the task of finding the source to enhance is delegated to the U-net module (which has access to the audio signal).
%We argue that in order to obtain optimal performance, the conditioning task should be performed in the conditioning module to let the U-net focus on the drift regression.
However, the source to enhance is typically the dominant concept shared between the visual scene and the audio. Since the conditioning module in VisAH only accesses visual information, the burden of identifying the target source is largely shifted to the U-Net, which has access to the audio signal. This forces the latter to implicitly perform source selection in addition to the regression task. To better disentangle these roles, we aim for the conditioning module to determine the source to enhance, leaving the U-Net to specialize in vector field estimation. %(as we use it in a flow matching setup).

%Inspired by the recent success of cross-modal adapters~\cite{CMT,crome}, we propose to use an adapter layer that would incorporate features from an additional modality into the image encoder (CLIP). Specifically, we propose to incorporate textual and audio features inside a CLIP layer by computing the cross-attention between projected (in lower dimension) CLIP and text/audio features and then re-projecting them to the original CLIP dimension. Formally, we define:
Inspired by recent cross-modal adapter designs~\cite{CMT,crome}, we propose an adapter layer that injects audio features from an additional audio encoder into intermediate layers of the CLIP model. 
The goal of this design is to let the CLIP representations dynamically attend to complementary audio cues, enriching the visual embeddings with audio-specific context. Specifically, we compute cross-attention between a low-dimensional projection of both the CLIP and audio features, before mapping them back to the original embedding dimension:
\begin{equation}
    \text{m}(F_k,E)=\mathrm{Att}(EW_{\text{down}}^{E},F_kW_{\text{down}}^{F
    })W_{\text{up}}^{E},
\end{equation}
%\HM{Note that the projection layers are not shared, as the input spaces of the encoders differ significantly in structure and scale.}
\iffalse
\begin{wrapfigure}{r}{0.60\linewidth}
    \centering
    \includegraphics[width=\linewidth]{fullVisahFM.png}
\caption{VisAH-FM builds on VisAH~\cite{visah} (purple), adding timestep conditioning, rollout loss, and an audio-aware cross-attention adapter inside CLIP (green).}
\label{fig:condDetailed}
    \vspace{-10pt}
\end{wrapfigure}
\fi
%Where $F_k$ corresponds to the original CLIP features at the layer $k$, $E$ the features for the other modality, $W_{down}^E$ and $W_{down}^{F_E}$ down projection layers, $W_{up}^E$ the up projection layer, and $Att$ the attention operation.
where $F_k$ denotes the CLIP features at layer $k$, $E$ the audio features, $W^{E}_{\text{down}}$ and $W^{F}_{\text{down}}$ the projection layers, $W^{E}_{\text{up}}$ the up-projection layer, and $\mathrm{Att}$ denotes cross-attention.
This operation effectively acts as a lightweight conditioning mechanism, enabling information exchange across modalities without retraining the CLIP backbone.
\begin{figure}
    \centering
    \includegraphics[width=0.8\linewidth]{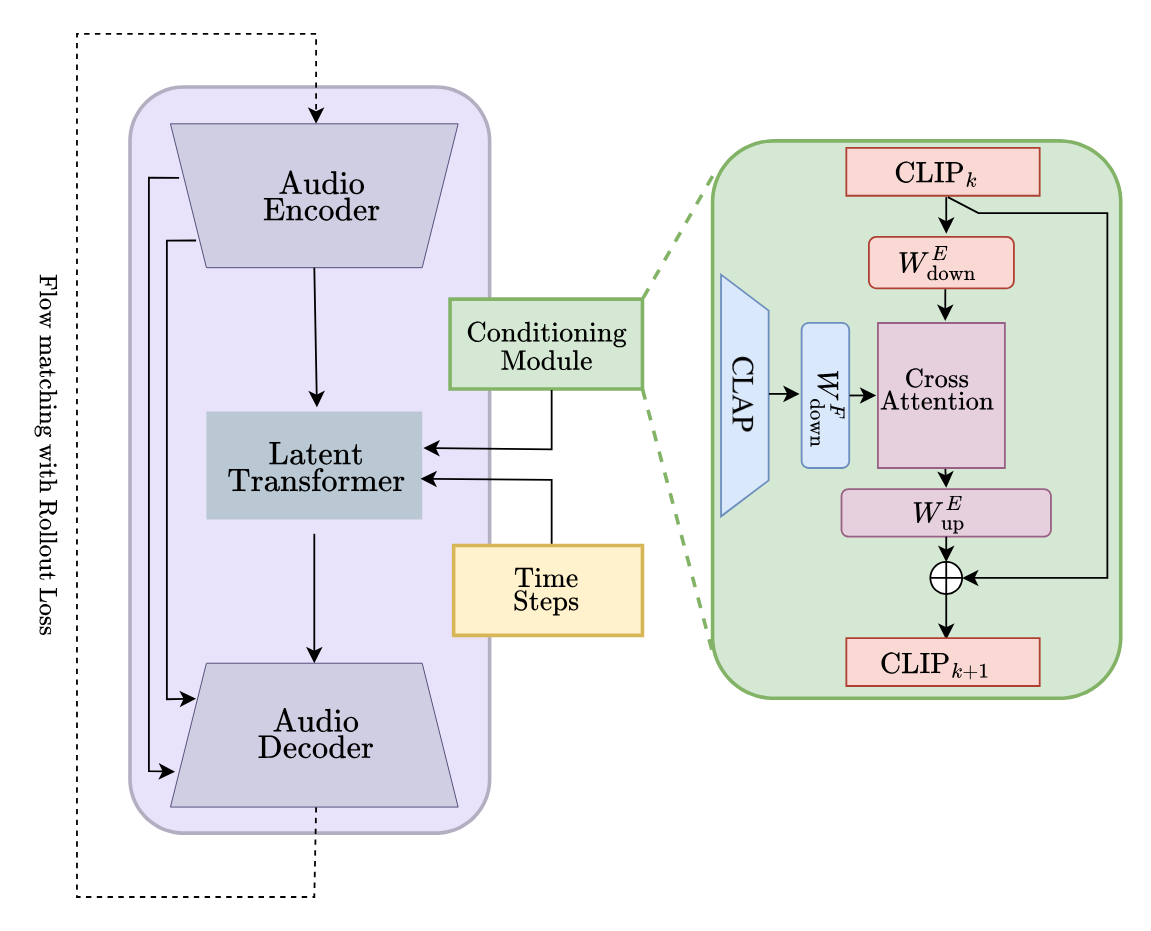}
    \caption{VisAH-FM builds on VisAH~\cite{visah} (purple), adding timestep conditioning, rollout loss, and an audio-aware cross-attention adapter inside CLIP (green).}
    \label{fig:condDetailed}
\end{figure}
%The audio is incorporated in the same manner, but using the audio features $A$ instead of the textual $E$. \\
%Note that, unlike in CMT\cite{CMT} we cant do at multiple layers of T5 XXL because too expensive \\
Finally, we perform a weighted sum of the original CLIP features with the audio-aware adapted features using learnable weights and pass this onto the next layer $k+1$. % $\lambda_E$ and $\lambda_A$:
 Formally, we define the output of the adapter layer as:\\
\begin{equation}
    \text{adapter}(F_k,E,A)=F_k + \lambda_E\text{m}(F_k,E) 
\end{equation}
where $\lambda_E$ is a learnable scalar controlling the contribution of the audio-aware features $E$. The overall conditioning mechanism is illustrated in Figure~\ref{fig:condDetailed} (green box).
Note that this conditioning can be applied to incorporate text features instead of, or along with, the audio.
\iffalse
\begin{figure}
    \centering
    \includegraphics[width=0.85\linewidth]{sec/condFull2.png}
    \caption{Proposed conditioning module: both the CLIP and the additional modality are projected in a lower-dimensional space where cross attention is performed. Features are then projected back in the original CLIP embedding space and summed with the original features.}
    \label{fig:condModule}
\end{figure}
\fi 
Importantly, initializing $\lambda_E$ to zero ensures that the first forward pass matches the pretrained CLIP encoder, allowing us to fully leverage pretrained weights while gradually incorporating cross-modal cues during training. In practice, we use $E$, the audio features of a CLAP \cite{CLAP} encoder as we expect that its alignment with text would simplify the incorporation to CLIP features.
%Note that when initializing $\lambda_E$ and $\lambda_A$ to 0, the first forward of the training would give the exact same original conditioning as the pre trained model, allowing to take maximally advantage of the pretrained model.
%\vspace{-1em}
\paragraph{Time (CFM) Conditioning.} 
We encode the flow matching timestep using sinusoidal embeddings, relying on the positional encoding scheme as adopted for diffusion timesteps in \cite{ddpm}, and append this embedding to the conditioning tokens provided to the latent transformer block via cross-attention. We ablate this choice in Appendix.

\section{Experiments}
\label{sec:exp}

% \subsection{Setting}
% \label{setting}
\paragraph{Implementation Details.}
%To optimally leverage the pretrained model, which was trained to directly predict well-balanced audio rather than a vector field, we systematically incorporate the source sample ($x_0$) into the model output so that it estimates a vector field instead of a \HM{balanced} audio. Specifically, we define $v_{\theta}(x_t,t,c) = x_0 - u_{\theta}(x_t,t,c)$, where $u_{\theta}(x_t,t,c)$ denotes the actual model.

To optimally leverage the pretrained VisAH model, which was trained to directly predict well-balanced audio rather than a vector field, we systematically incorporate the source sample ($x_0$) into the model output to define the estimated vector field as
$v_{\theta}(x_t,t,c) = x_0 - u_{\theta}(x_t,t,c)$,
where $u_{\theta}(x_t,t,c)$ denotes the actual model.
We trained the model for 50 epochs ($\sim$23,500 iterations) using a learning rate of $1 \times 10^{-4}$, a cosine annealing scheduler, and a batch size of 32.\\

Following VisAH, the input waveform was sampled at 44.1\,kHz and converted to mono by averaging the stereo channels. We observed an issue with audio saving quality, and therefore resampled the test set using the official code from \cite{visah}. We also trained another model on the exact same dataset as \cite{visah}, without on-the-fly remixing, and present the results in Appendix. 
Similar to \cite{LBM}, we discretized time into four timesteps, sampled them uniformly during training, and used fixed-step Euler integration at inference. 
Although our model performs several recurrent passes of the vector-field estimator at inference, the resulting overhead is minor. Most computation is dominated by the conditioning encoders — CLIP ($\sim$300 M parameters), CLAP ($\sim$31 M), and optionally T5-XXL (11B) or InternVL (8B) - which are run once per video segment, whereas the repeated U-Net vector-field estimator is comparatively lightweight ($\sim$60M). The adapter module is incorporated at the $18^{\text{th}}$ layer of the CLIP model, and training-time VRAM measurements are reported in Appendix.
Except for the changes mentioned in Sec.~\ref{sec:method}, we used the exact VisAH architecture.  All experiments were performed on the Muddy Mix Dataset. Additional details are given in Appendix.%, (as it is the only one available for the task of visually guided acoustic highlighting).
%% \vspace{-1em}
\paragraph{Metrics.}
We evaluate our models using the same metrics as VisAH, along with the introduction of an additional remixing metric:
\begin{itemize}
    \item \textbf{Signal metrics:} We measure the magnitude distance (Mag)~\cite{mag}, which evaluates audio quality in the time–frequency domain, and envelope distance (Env)~\cite{env}, which assesses quality in the time domain. We also measure temporal alignment using the Wasserstein distance (Was).
    \item \textbf{Semantic alignment:} We compute the KL divergence (KLD) between the predicted logits of a PASST~\cite{PASST} model for the predicted and ground-truth audio. Additionally, we report the ImageBind score (IB Score)~\cite{IB} to quantify the similarity between predicted audio and image frames. Unlike VisAH, we report the IB Score directly instead of $\Delta \text{IB}$, as we find it more interpretable.
    \item \textbf{Remixing metric:} Since the goal is to adjust the relative loudness across three categories—human speech, music, and sound effects—we employ a sound source separation model~\cite{bandsplit} to extract stems from both the predicted and ground-truth audio, and measure the loudness difference (LDif) between corresponding sources as $\frac{1}{K} \sum_k \big| \mathrm{loud}(s_k(\hat{x})) - \mathrm{loud}(s_k(x)) \big|$,
    % \begin{equation}
    %     L_\mathrm{Dif} = \frac{1}{K} \sum_k \big| \mathrm{loud}(s_k(\hat{x})) - \mathrm{loud}(s_k(x)) \big|,
    % \end{equation}
    where $\hat{x}$ and $x$ denote the predicted and ground-truth audio, respectively, $\mathrm{loud}(.)$ computes the loudness, and $s_k(.)$ the separator of the $k$-th source from the audio.
\end{itemize}

%We also introduce a new metric to compute the remixing distance with the ground truth mix. By forwarding a sound source separation model on the predicted and ground truth audio to separate it in three tracks: speech, sound effects and music

\subsection{Quantitative Analysis}
\begin{table*}
    \caption{Main results. Best results are highlighted in \textbf{bold}, and all values except the LDif are multiplied by 100.}
    \setlength{\tabcolsep}{5.5pt}
    \renewcommand{\arraystretch}{1.18}

    \scriptsize
    \centering
    \begin{tabular}{llcccccc}
        \toprule
        Model & Conditioning & IB Score $\uparrow$& KLD $\downarrow$& LDif $\downarrow$& Mag $\downarrow$& Env $\downarrow$& Was $\downarrow$\\
        \midrule
        Input & - & 28.14 & 20.74& 18.36& 22.69&6.29 & 1.96 \\
        \midrule
        VisAH &CLIP & 28.84 & 11.37 & 9.66 & 9.99 & 3.38 & 0.84 \\
        VisAH &T5 & 28.92 & 11.71 & 9.63 & 10.22 & 3.44 & 0.88 \\
        VisAH & CLIP-CLAP & 28.82 & 11.28 & 9.60 & 10.27 & 3.56 & 0.80 \\
        VisAH-FM (Ours) & CLIP-CLAP & \textbf{29.12} & \textbf{9.70} & \textbf{7.77} & \textbf{8.28} & \textbf{2.74} & \textbf{0.63} \\
        \bottomrule
    \end{tabular}
     %\vspace{7pt}
    
    \label{tab:main_table}
\end{table*}
\subsubsection{Main Results.}
Table~\ref{tab:main_table} compares our proposed model VisAH-FM against the VisAH baseline using image (CLIP), text (T5-encoded), or our audio-adapted CLIP-CLAP conditioning. Adding the adapter conditioning to the discriminative VisAH baseline slightly improves performance, showing that the conditioning module is independently useful. However, VisAH-FM still clearly outperforms all VisAH variants, indicating that the gains come from both the improved conditioning and the flow matching formulation.
%Table~\ref{tab:conditioning} present the performance of the flow matching model using different types of conditioning. Interestingly, without CMT adapters, the vision conditioning seems to perform slightly better than the textual conditioning. Incorporating additional modality in the image encoder boost performance consistently in all the metrics, however, the addition of the audio modality (encoded through CLAP~\cite{CLAP}) boost significantly more the results than the textual features. 
% \vspace{-0.5em}
\paragraph{Ablation on Conditioning Module.} Table~\ref{tab:conditioning} presents an ablation study of the different conditioning modalities and methods for the flow matching model. Without adapters, vision-based conditioning performs slightly better than text-based conditioning. Incorporating additional modalities into the image encoder consistently improves performance across all metrics. Notably, adding audio features encoded with CLAP~\cite{CLAP} yields larger gains than adding textual features.
%We attribute this behavior to the audio-visual interaction happening in the conditioning module: the source to enhance is being determined in the conditioning module to let the drift estimator module focus on the drift regression. The text features encodes the same kind of information as the image features (i.e., the visual scene), on the other hand, the audio features encode the auditory scene. As the source to enhance generally corresponds to common concept between the auditive and visual scene, then addition of the audio modality is more beneficial than the text one.\\
We attribute this improvement to the stronger audio–visual interaction enabled by the conditioning module: the module determines which source to enhance, allowing the velocity-field estimator to focus only on regression. As text captions originate from a VLM, they mainly contain information similar to image features (i.e., visual scene semantics), whereas audio features encode the auditory scene. Since the target source typically corresponds to elements common to both the visual and auditory domains, integrating audio features inside the conditioning proves more beneficial than adding the textual counterpart.
%Additionally, CLAP might extract different audio features than the DEMUCS architecture which can be beneficial to better understand the auditory scene.
Additionally, CLAP extracted audio representations are complementary to those from the DEMUCS encoder, further enriching the model’s understanding of the auditory scene.
\iffalse
\begin{table}[!t]
    \centering
    \begin{tabular}{cccc}
        \toprule
        Model & IB Score $\uparrow$& KLD $\downarrow$& LDif $\downarrow$\\
        \midrule
        FM CMT Audio+Text & 29.10 & 9.73 & 7.90 \\
        FM CMT Audio & \textbf{29.12} & \textbf{9.70} & \textbf{7.77} \\
        FM CMT Text & 29.08 & 9.76 &7.85   \\
        FM Vision & 29.09 & 9.79 & 7.87  \\
        FM Text & 28.88 & 10.50 & 8.13 \\
        \bottomrule
    \end{tabular}
    \caption{Ablation on the conditioning. We compare different conditioning methods (with and without adapter) and different modalities (audio, image, and text).}
    \label{tab:conditioning}
\end{table}
\fi
\begin{table}[!t]
    \setlength{\tabcolsep}{10pt}
    \renewcommand{\arraystretch}{1.12}
    \caption{Conditioning ablation. We compare different conditioning methods (with and without adapter) and different modalities: audio (A), vision (V), and text (T).}
    \label{tab:conditioning}
    \centering
    \footnotesize
    \begin{tabular}{ccccc}
        \toprule
        Adapter & Modality & IB Score $\uparrow$& KLD $\downarrow$& LDif $\downarrow$\\
        \midrule
        \checkmark & V+A& \textbf{29.12} & \textbf{9.70} & \textbf{7.77} \\
        \checkmark & V + T & 29.08 & 9.76 &7.85   \\
        \checkmark & V+T+A &29.10 & 9.73 & 7.90 \\
        \ding{55} & V & 29.09 & 9.79 & 7.87  \\
        \ding{55} & T &28.88 & 10.50 & 8.13 \\
        \bottomrule
    \end{tabular}
\end{table}
%% \vspace{-1em}
%Interestingly, incorporating both the textual and audio features slightly perform slightly worse than simply using visual and audio modality. %We hypothesize that 
Interestingly, incorporating both textual and audio features performs slightly worse than using only visual and audio modalities, indicating that textual features are not needed when the audio is directly incorporated into the image encoder. This finding holds practical importance for real-world applications as the extraction of text features is very costly, requiring a forward pass through a VLM~\cite{internVL} and a large text encoder~\cite{T5}.%This may indicate mild redundancy between text and audio cues, suggesting that the audio–visual pair provides a more focused conditioning signal.
%Table~\ref{tab:comparison_rollout} ablate the rollout loss and compare it with some standard methods used to improve consistency across steps of flow matching model. The addition of the rollout loss plays a crucial role in the performance of the flow matching model, as removing it drops significantly the performances ($\sim$1.2 of KLD and $\sim$1.6 of loudness difference). 
%% \vspace{-1em}
\paragraph{Ablation of the Rollout Loss.} Table~\ref{tab:comparison_rollout} ablates the rollout loss and compares it with several standard methods designed to improve consistency across the flow matching steps. The addition of the rollout loss proves crucial: removing it leads to a marked performance drop \textit{i.e.} approximately +1.2 in KLD and +1.6 in loudness difference.

%Consistency loss forces the vector field estimated at time $t$ to be the same as the one estimated at time $t+\Delta$, starting from the previous estimated step. More formally it is defined as:
The \textbf{consistency loss} encourages the vector field predicted at time $t$ to remain consistent with the one predicted at a slightly later time $t+\Delta$, when the input is advanced along the flow using the earlier prediction:
\begin{equation}
    \mathcal{L}_{\text{cons}}=\|v_{\theta}(x_t,t)-\text{StopGrad}(v_{\theta}(x_t - \Delta_tv_{\theta}(x_t,t),t+ \Delta_t))\|
\end{equation}
%Interestingly, adding this loss to the model significantly degrades its performance. We hypothesize that this is due to the fact that if the first estimate is noisy, the second one would be even worse, hence guiding the flow matching process with a wrong direction. On the other hand, the rollout loss anchors the prediction to the ground truth itself, avoiding this trajectory degeneration.
Interestingly, adding this loss substantially degrades performance. We hypothesize that this occurs because if the first estimate is noisy, the propagated estimate becomes even less accurate, effectively guiding the model in an incorrect direction. In contrast, the rollout loss anchors predictions to the ground-truth trajectory, preventing such drift and stabilizing training.
%Recent works underlined the efficiency of adding gaussian noise to the input trajectories \cite{LBM,stochasticInterpolant}, making the model more robust to out-of-distribution trajectories at inference time. The line 'FM with noise' present the results obtained when training a model with noise added to the input trajectories. While it slightly outperform the regular flow matching, this method falls short compared to the model trained with rollout loss.
%In scenarios where the model rely on a VAE, the input space is supposed gaussian, hence the noise can mimic realistic errors, however in our case, the input space is the raw audio, hence the noise is corresponds to white noise, which is significantly different from the error that the model might produce. 
%Hence, we argue that the small perturbation introduced in the trajectories fails at imitating model's error, and therefore at making the model robust to it.   
%The rollout loss significantly outperforms it as it uses the exact output of the model as the input, replicating perfctly the inference scenario.\\

Recent works have shown that injecting Gaussian noise into the input trajectories can improve robustness to out-of-distribution trajectories at inference time~\cite{LBM,bridgeMatching}. The row \textit{``Bridge Matching''} reports the results of a model trained with this strategy (using Gaussian noise centered on zero with standard deviation of 1e-5). While it slightly outperforms the standard flow matching baseline, it remains notably inferior to the version trained with rollout loss. In VAE-based setups, the latent space is assumed Gaussian, hence, added noise can mimic realistic errors. However, in our case, the input space corresponds to raw audio, and the injected noise effectively acts as white noise—significantly different from the model’s own prediction errors. Consequently, these perturbations fail to simulate realistic inference conditions, and robustness improvements remain limited. In contrast, the rollout loss directly reuses the model outputs as inputs, accurately reproducing the inference scenario and yielding significantly better performance.
%Finally, the line 'FM with t weighting' shows the result of a flow matching model trained with a loss that weights error proportionally to the time value used to generate the input (i.e., put higher weights to samples close to the input distribution). This baseline performs slightly worse than the regular flow matching method, showing that forcing the model to focus on early steps is not enough, small errors are inevitable and the forcing the model to be able to correct its own errors across time is crucial, highlighting the effectiveness of the rollout loss.

The row \textit{``FM Weighted''} corresponds to a flow matching model trained with a loss weighted inversely proportional to the time variable, emphasizing samples closer to the input distribution. This baseline performs slightly worse than standard flow matching, indicating that simply prioritizing early steps is insufficient. Instead, enabling the model to iteratively correct its  errors across time proves essential—highlighting the effectiveness of the rollout loss.
\begin{table}[t]
\centering

\begin{minipage}[t]{0.49\linewidth}
    \caption{Rollout component ablation.}
    \label{tab:comparison_rollout}
    \centering
    \renewcommand{\arraystretch}{1.12}
    \scriptsize
    \begin{tabular}{lccc}
    \toprule
         Objective & IB Score $\uparrow$ & KLD $\downarrow$ & LDif $\downarrow$\\
         \toprule
         FM + Rollout& \textbf{29.09} &\textbf{9.79}  & 7.87 \\
         \midrule
         FM& 28.92 & 10.99 & 9.48\\
         Rollout & 28.94 & 9.92 & \textbf{7.71}\\
         FM + Consistency & 28.40 & 14.58 & 10.30  \\
         Bridge Matching & 29.04 & 10.85 & 9.62  \\
         FM Weighted & 28.86 & 11.16 & 9.44 \\
         \bottomrule
    \end{tabular}    
\end{minipage}
\hfill
\begin{minipage}[t]{0.49\linewidth}
    \caption{Hyperparameter sensitivity.}
    \label{tab:HP}
    \renewcommand{\arraystretch}{1.12}
    \scriptsize
    \centering
    \begin{tabular}{cccccc}
    \toprule
         $\lambda$ & Horizon & Steps & IB Score $\uparrow$ & KLD $\downarrow$ & LDif $\downarrow$\\
         \toprule
         0.1& 4 & 4& 29.12 & 9.85 & 8.02  \\
         0.3& 4&4&29.09 &9.79  & 7.87 \\
         0.5& 4&4&29.12 & 9.93 & 7.80  \\
         1.0& 4&4&29.04 & 9.82 & 7.84  \\
         \midrule
         0.1 & 2&4&29.08 & 9.68 & 7.94  \\
         0.3& 4&10&29.10 & 9.79 & 8.35  \\
         \bottomrule
    \end{tabular}    
\end{minipage}
\end{table}
The row \textit{``Rollout''} corresponds to a model trained without the standard flow matching loss, relying solely on the rollout loss. Remarkably, this variant noticeably outperforms vanilla flow matching and only slightly underperforms the full model that combines flow matching and rollout loss. This highlights the central role of the rollout mechanism: even without the intermediate supervision and linearity prior imposed by flow matching, the model achieves strong performance.
Moreover, this variant significantly surpasses the original VisAH model, suggesting that allowing the model to repeatedly refine its predictions through multiple forward passes over the same architecture yields meaningful improvements. This iterative refinement behavior is evocative of phenomena observed in large language models, where multi-step reasoning improves output quality (e.g., chain-of-thought prompting~\cite{wei2022chain,wang2022self}). We further analyze the behavior of this model and its learned trajectories in Section~\ref{sec:behave}.
%Table~\ref{tab:HP} presents the sensitivity of the model with respect to the hyper parameter of the rollout loss. 
% \vspace{-1em}
\paragraph{Hyperparameter Sensitivity.} Table~\ref{tab:HP} reports the sensitivity of our method to hyperparameters of the rollout loss. 
%The first lines indicate the coefficient of the rollout loss in the final loss (i.e. the value of $\lambda$ in Eq.~\ref{eq:finalLoss}). The performance remains quite stable, showing the resilience of the method to this hyperparameter.
The first block of the table varies the rollout coefficient $\lambda$ (Eq.~\ref{eq:finalLoss}). Performance remains stable across a broad range, highlighting the robustness of the method to this hyperparameter.
%We variate the horizon of the rollout: the number of steps performed before applying the rollout loss (the starting point is chosen randomly when the horizon is smaller than the total number of steps). The last two line shows experiments where the horizon is smaller than the total number of inference steps. The performance do not decrease significantly, underlining the behavior of the rollout loss: not all the steps are needed, performing a few steps is enough to train the model to be resilient to its own previous error.
We also vary the rollout horizon, defined as the number of steps before applying the rollout loss. When the horizon is shorter than the full inference trajectory, we randomly choose the starting step to ensure coverage of different temporal segments; otherwise, rollout begins from the first step. The last two rows correspond to settings where the rollout horizon is smaller than the total number of inference steps. We observe no significant performance degradation, suggesting that rollout over only a subset of steps is sufficient to teach the model to recover from its own intermediate errors.%—full unrolling is not required during training.
% \vspace{-1em}
\begin{figure*}[h]
    \centering

    \begin{subfigure}[t]{0.53\linewidth}
        \centering
        \includegraphics[width=\linewidth]{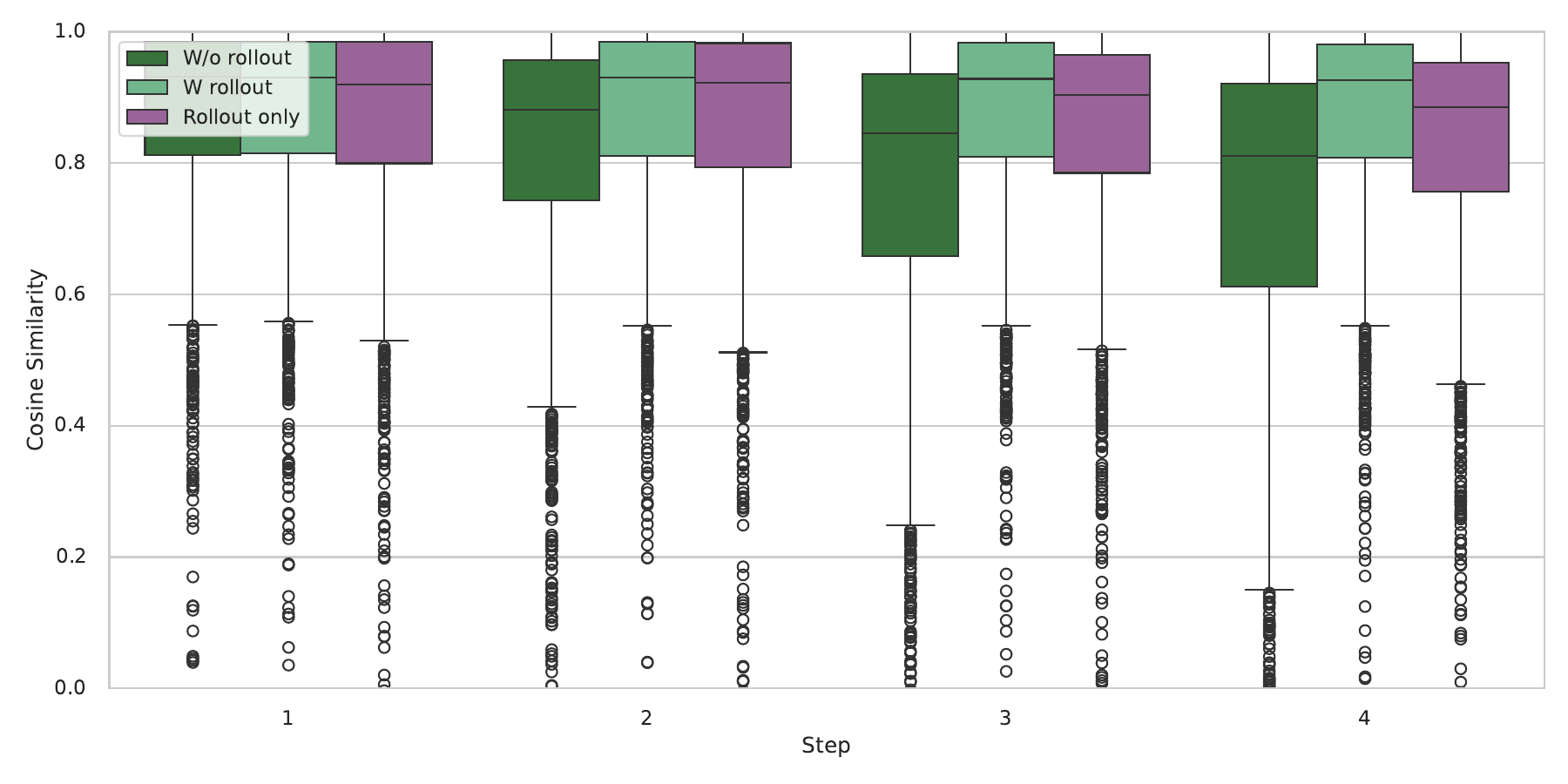}
        \caption{Evolution of the cosine distance between predicted and ground truth trajectories across steps.}
        \label{fig:cosinePlot}
    \end{subfigure}
    \hfill
    \begin{subfigure}[t]{0.40\linewidth}%{0.29\linewidth}
        \centering
        \includegraphics[width=\linewidth]{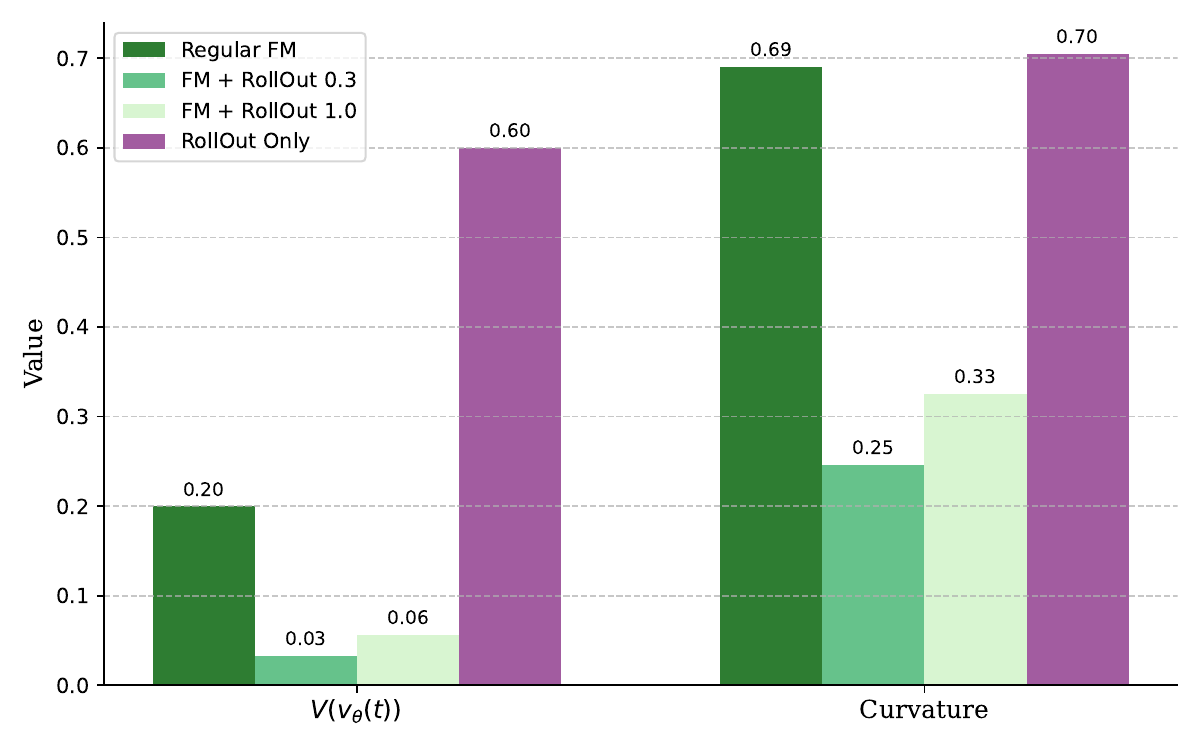}
        \caption{Variance, average curvature, and tortuosity of the predicted vector fields w/ and w/o rollout loss.}
        \label{fig:linearity}
    \end{subfigure}
    %\hfill
    %\begin{subfigure}[t]{0.28\linewidth}
    %    \centering
    %    \includegraphics[width=\linewidth]{sec/trajPasst.pdf}
    %    \caption{Trajectories with and without rollout loss, in the PASST space.}
    %    \label{fig:trajPasst}
    %\end{subfigure}

    \caption{Impact of rollout loss: with flow matching it stabilizes trajectories and limits error accumulation; alone it learns nonlinear paths.}
    \label{fig:three_figs}
\end{figure*}
\subsubsection{Model Behavior.}
\label{sec:behave}
% \vspace{-1em}
Here we analyze how the rollout loss shapes the learned trajectories, and how these differ from those obtained with standard flow matching.
% \vspace{-1em}
\paragraph{Error Accumulation Analysis.}
%Figure~\ref{fig:cosinePlot} shows the cosine similarity between the estimated and ground truth vector field across the test set, through the steps of the inference process. 
%As hypothesized previously, the model trained without rollout loss goes further from the ground truth as the steps go: as the first estimate is imperfect, the model goes on an out-of-distribution trajectory and worsens its predictions across steps. On the other hand, the model trained jointly with rollout and flow matching loss stays close to the ground truth trajectory across all steps. Interestingly, the model trained with rollout loss only slightly goes away from the ground truth trajectories across steps. 
Figure~\ref{fig:cosinePlot} reports the cosine similarity between the predicted and ground-truth vector fields over inference steps on the test set. Consistent with our hypothesis, the model trained without rollout loss progressively diverges from the ground-truth trajectory: initial errors compound and drive the model toward out-of-distribution regions, leading to worsening predictions over time. In contrast, the model trained with both rollout and flow matching losses maintains high alignment throughout the trajectory, demonstrating that the rollout objective effectively constrains error accumulation and stabilizes long-range trajectories. Interestingly, the model trained with rollout loss alone exhibits small drift during the last steps, although it performs well overall (see Table~\ref{tab:comparison_rollout}), indicating that the learned trajectories differ from the linear one imposed in flow matching.
% \vspace{-1em}
\paragraph{Trajectory Linearity.}
%In order to evaluate how linear the predicted trajectories are, we computed the the average variance of the predicted vector field across steps, as well as the average discrete curvature, on the test set.
Next, we study how the rollout loss affects the geometry of the learned trajectories. In flow matching, the optimal trajectory is theoretically linear in the data space; however, finite-capacity models and error accumulation during inference can induce nonlinear deviations. To quantify this effect, we compute both the variance of the predicted vector field across inference steps and the average discrete curvature of the trajectory on the test set.
Formally, we define the discrete curvature at step $t$ as
\begin{equation}
    \Theta_t = \arccos\!\left(\frac{\hat{v}_{t+1} \cdot \hat{v}_t}{\|\hat{v}_{t+1}\| \,\|\hat{v}_t\| + \epsilon}\right),
\end{equation}
where $\hat{v}_t$ denotes the predicted vector field at time $t$. $\Theta_t$ measures the angular deviation between successive vector field directions; higher values indicate stronger trajectory bending. Similarly, higher variance across steps reflects instability in the predicted vector fields.
%Where $\hat{v}_t$ is the estimated vector field by the model at time $t$. $\Theta_t$ measures the angle between the vector field predicted at time $t$ and $t+1$.
%Intuitively, the higher the variance of the vector fields and the curvature, the more non-linear the trajectories are.
%Figure~\ref{fig:linearity} shows the average variance of the predicted vector fields as well as the average discrete curvature for diffrent model trained with: flow matching only, flow matching and rollout loss with coefficient of 0.3 and 1.0 and rollout solely. As the model trained only with flow matching loss falls into out-of-distribution trajectories after one step, its trajectories are highly non-linear. Interrestingly, adding the rollout loss stabilizes the trajectories and make them much more linear. However, as the coefficient given to the rollout loss increases the generated trajectories are less and less linear. The model trained with rollout loss only generates highly non-linear trajectories, explaining the behavior observed in Figure~\ref{fig:cosinePlot}: the vector fields are not aligned with the ground truth one as the model is taking a completely different (non-linear) path to reach the target.\\
 Figure~\ref{fig:linearity} reports average curvature and variance for models trained with (i) flow matching only, (ii) flow matching with rollout loss ($\lambda \!=\!0.3$ and $1.0$), (iii) rollout loss only. The flow matching-only model exhibits strong nonlinear deviations after the first step, consistent with error compounding and off-manifold drift. Introducing the rollout loss markedly stabilizes the trajectory, yielding substantially lower curvature and variance. 
Interestingly, when the rollout term increases, trajectories become less linear again.
Consequently, when using rollout loss alone, the model learns highly non-linear trajectories, explaining the behavior observed in Figure~\ref{fig:cosinePlot}: predicted vector fields are not aligned with the ground-truth, as the model takes a completely different path to reach the target.
%Figure~\ref{fig:trajPasst} shows a PCA of trajectories (embedded in the PASST\cite{PASST} space) of the model trained with flow matching with and without rollout loss, as well as the associated ground truth. While the models perform similarly at the first step, the difference becomes significant after the second step where the model trained without rollout starts taking very noisy trajectories while the other remains close to the ground truth. Additionally, in cases where the two models start in a bad direction, we still see the impact of the rollout loss: the end trajectory slightly corrects its previous error.
%Figure~\ref{fig:trajPasst} visualizes PCA-projected trajectories in the PASST~\cite{PASST} embedding space for models trained with and without the rollout loss, alongside the ground-truth trajectory. Both models follow a similar direction at the first step, but clear differences emerge by the second step: the flow matching-only model quickly deviates and produces noisy trajectories, whereas the rollout-trained model remains close to the ground truth. Moreover, even when both models initially move in an incorrect direction, the rollout-trained model partially corrects its path over subsequent steps, showing robustness to early prediction errors.
\subsection{Qualitative Analysis}
%FM allows slider for remixing which rollout only does not really

\begin{figure}[h]
    \centering
    \includegraphics[width=0.85\linewidth]{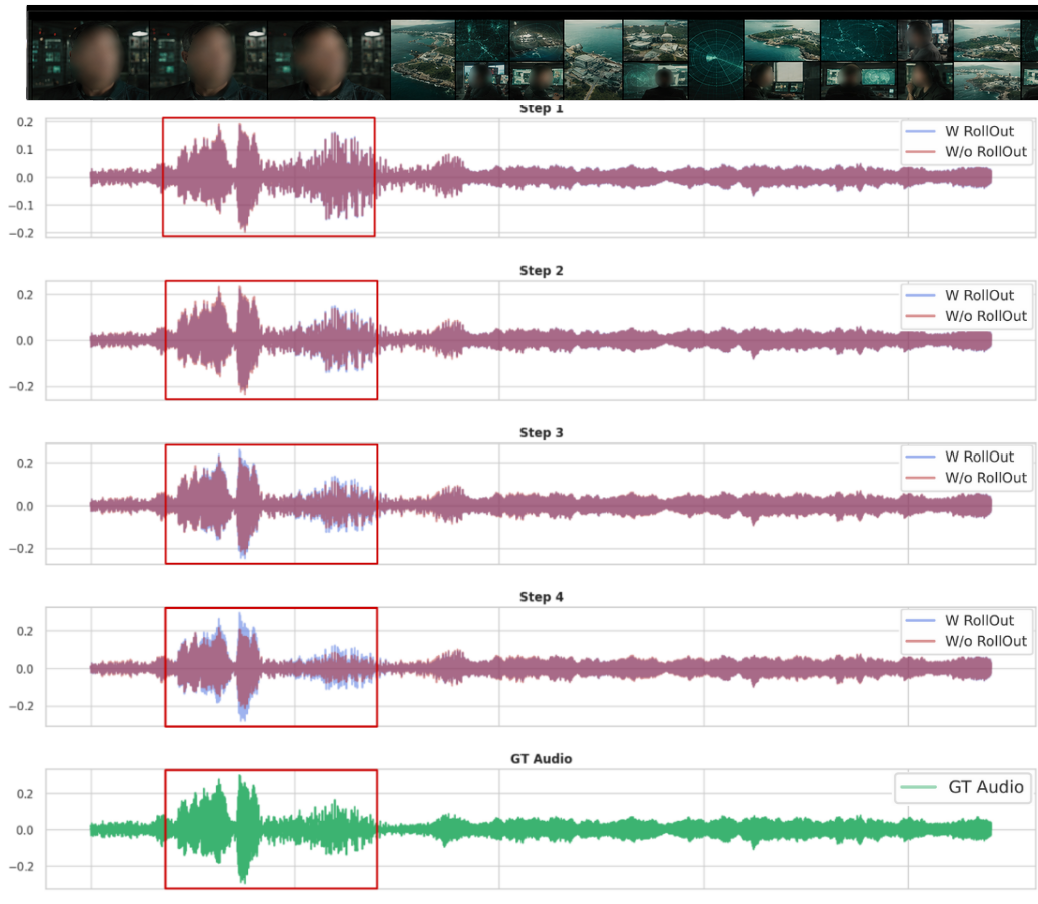}
    \caption{Illustration of the behavior of the model
trained with and without rollout loss. Wave-
forms are overlapped for readability. The őrst
step is similar, but the difference appears in the
latter steps when the rollout-trained model en-
hances speech, while the other does not.}
    \label{fig:qualiWavRoll}
\end{figure}

%% \vspace{-0.5em}
%\noindent\begin{minipage}[t]{0.48\linewidth}
\paragraph{Visualization of the Rollout Loss.}
Figure~\ref{fig:qualiWavRoll} shows outputs across flow-matching steps. The rollout-trained model reliably enhances speech (red boxes), whereas the regular flow model does not. Since each step remains realistic, the step count can act as a knob for highlighting strength.
Additionally, as the flow maps poorly-balanced audio to well-balanced audio (as opposed to starting from noise), the output of each step is incrementally transformed realistic audio.
This unlocks the possibility for a user to use the number of steps as a knob to specify the degree of desired highlighting.
%\end{minipage}\hfill
%\begin{minipage}[t]{0.48\linewidth}
%\centering
%\vspace{-0.9em}
%\includegraphics[width=0.78\linewidth]{fig5Blur.jpeg}
%\captionof{figure}{With rollout loss, later FM steps enhance speech more reliably; waveforms are overlaid.}
%\label{fig:qualiWavRoll}
%\end{minipage}
%\paragraph{Qualitative visualization of visahFM against VisAH} We compare the waveform obtained by our model to the remixing obtained using the discriminative model VisAH in Figure~\ref{fig:qualiFMVisah}.  The top left sample highlights where the VisAH model did not enahance enough the main source, but enhanced an event at the end of the audio. The flow matching model emphasized the first event and less the last one. The three other examples showcase samples where the VisAH model did not enhance are not enough the source to increase while the flow matching model did. 
%\paragraph{Visualization of \modelname{} against \taskname{}} Figure~\ref{fig:qualiFMVisah} presents qualitative waveform comparisons between our flow-matching model and the discriminative VisAH baseline. The left example illustrates a case where VisAH fails to sufficiently enhance the primary source and instead amplifies a secondary event near the end of the clip. In contrast, our model correctly emphasizes the dominant event at the beginning while enhancing less the latter one. However, one might notice that the source to decrease is not decreased enough, showing that visahFM reamains prone to error and further work is needed to keep improving it.
%The right-sided example further illustrates a case where VisAH does not adequately boost the target source, whereas VisAH-FM consistently produces stronger and more focused enhancement.
\paragraph{Visualization of \modelname{} against \taskname{}.}
\iffalse
\textcolor{blue}{EDIT PARAGRAPH}
In Figure~\ref{fig:qualiFMVisah}, top example shows a case where VisAH does not adequately boost the target source, whereas VisAH-FM consistently achieves stronger and more focused enhancement. The bottom example illustrates another case where VisAH fails to sufficiently enhance the primary source and instead amplifies a secondary event near the end of the clip. In contrast, our model correctly emphasizes the dominant event at the beginning while only mildly enhancing the latter one. Nevertheless, the undesired source is not fully suppressed, indicating that VisAH-FM still exhibits residual artifacts and leaves room for further improvement.
\fi
Figure~\ref{fig:qualiFMVisah} shows two cases where VisAH-FM does consistently better than VisAH when attenuating overly loud sources. Occasional artifacts in VisAH-FM output leave room for further improvement. More visualizations are available in the Appendix. Demo examples are provided in the supplements.
%However, although more attenuated the output of VisAH-FM remains slightly different than the original ground truth (sometimes showing small artifacts), leaving room for further improvement.

%The right-sided example illustrates another case where VisAH fails to sufficiently decrease the source near the end of the clip. In contrast, our model correctly emphasizes the dominant event at the beginning while only mildly enhancing the latter one. Nevertheless, the undesired source is not fully suppressed, indicating that VisAH-FM still exhibits residual artifacts and leaves room for further improvement.

\iffalse
\begin{figure}
    \centering
    \begin{subfigure}{0.32\textwidth}
        \centering
        \includegraphics[width=\linewidth]{plot20Flat.png}
    \end{subfigure}
    \hspace{0.02\textwidth}
    \begin{subfigure}{0.32\textwidth}
        \centering
        \includegraphics[width=\linewidth]{plot4Flat.png}
    \end{subfigure}

    \caption{Qualitative comparison between VisAH and VisAH-FM. Main differences are framed in red, and waveforms are overlapped for readability.}
    \label{fig:qualiFMVisah}
\end{figure}
\fi

\begin{figure}
\centering
\includegraphics[width=0.49\linewidth,height=1.2cm]{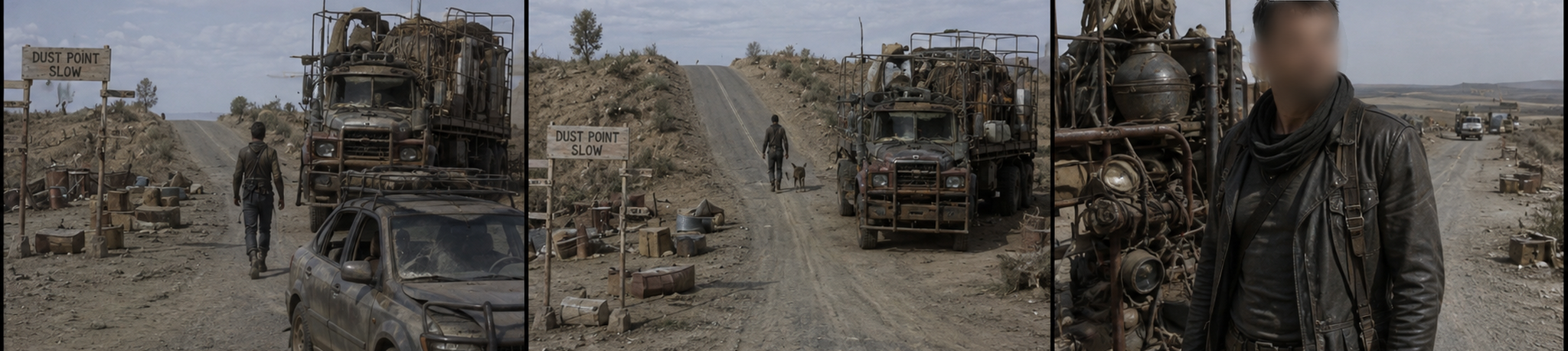}
\includegraphics[width=0.49\linewidth,height=1.2cm]{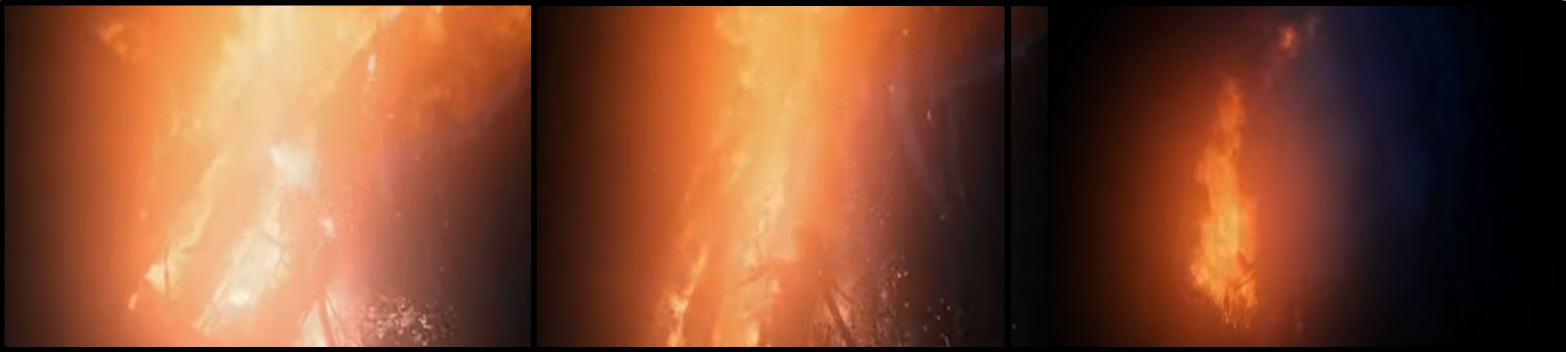}

\includegraphics[width=0.49\linewidth,height=1.4cm]{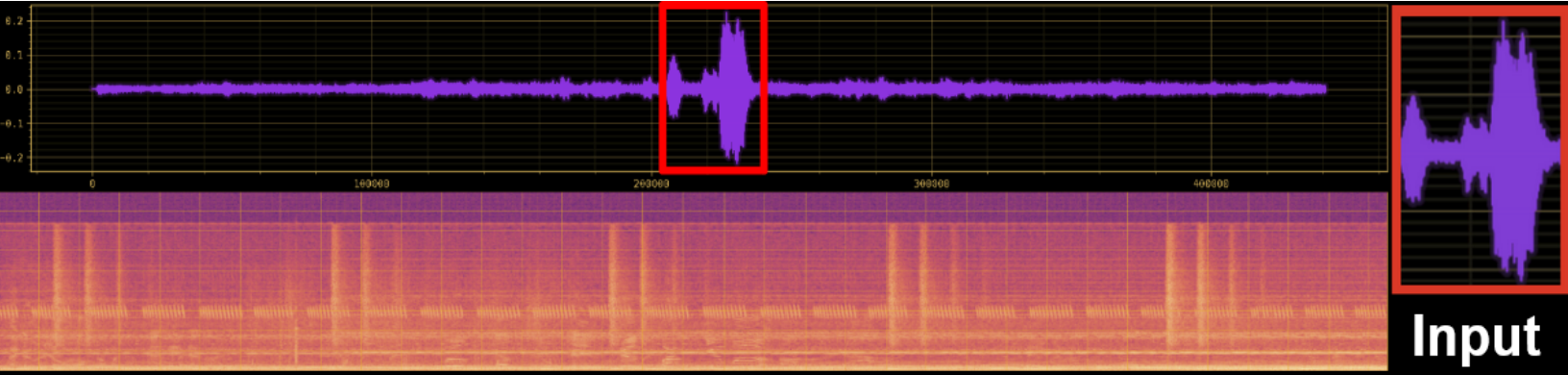}
\includegraphics[width=0.49\linewidth,height=1.4cm]{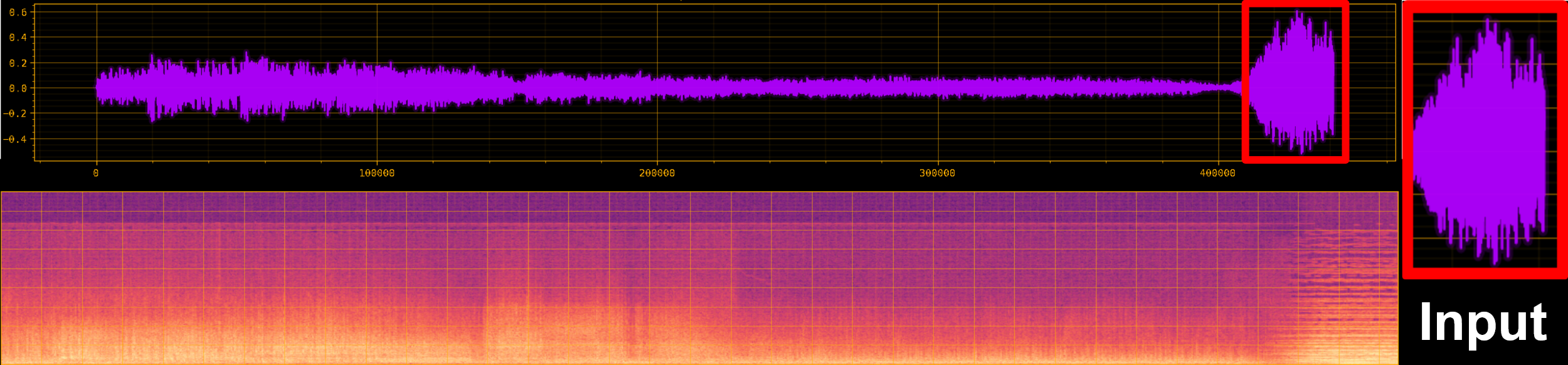}

\includegraphics[width=0.49\linewidth,height=1.4cm]{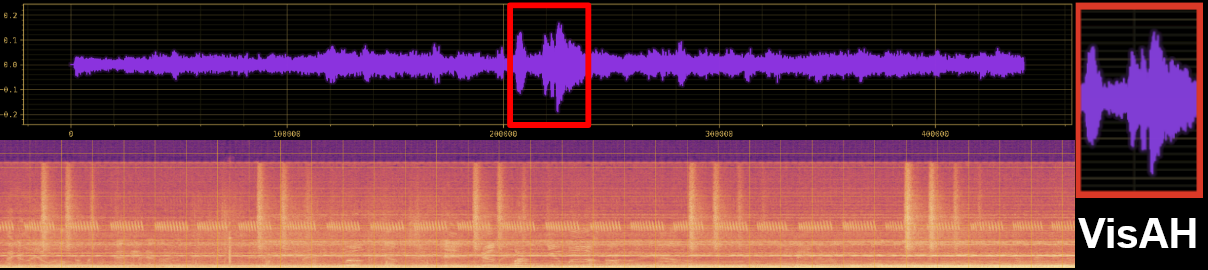}
\includegraphics[width=0.49\linewidth,height=1.4cm]{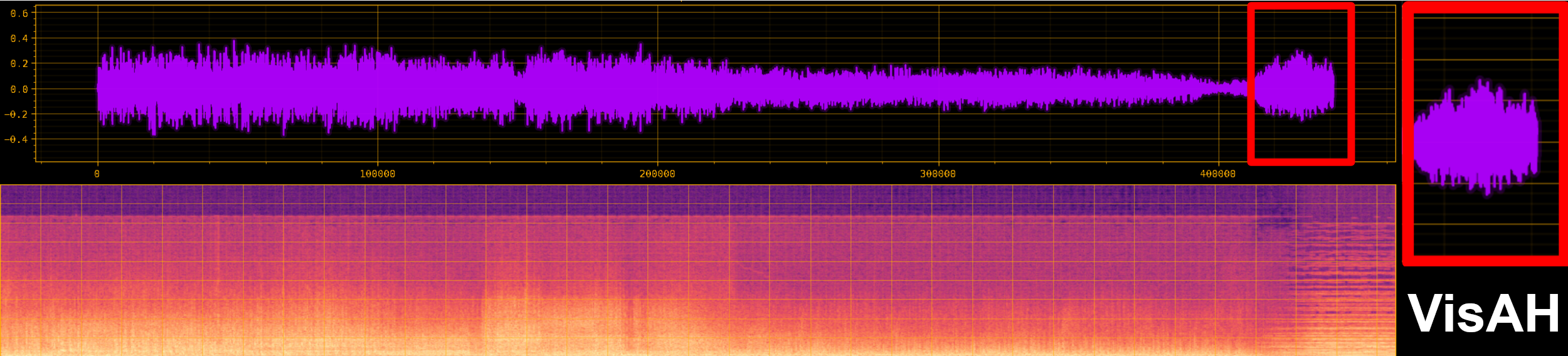}

\includegraphics[width=0.49\linewidth,height=1.4cm]{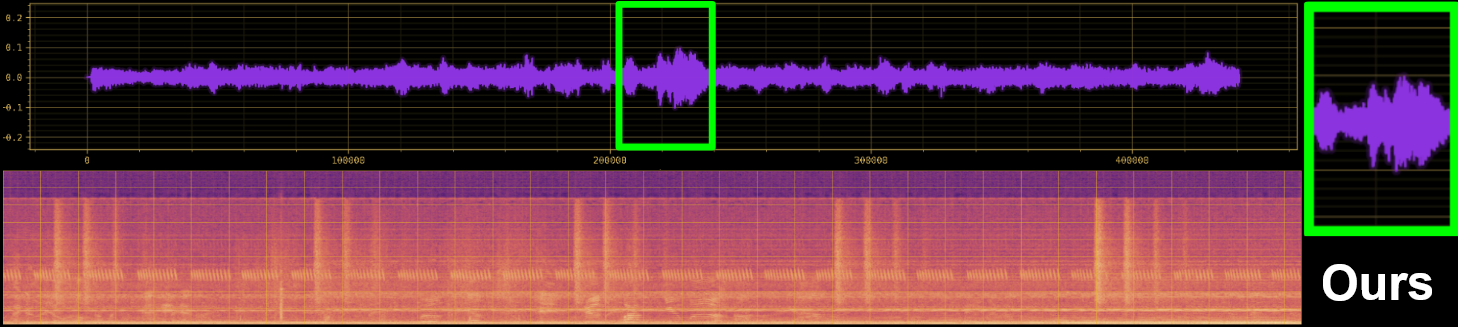}
\includegraphics[width=0.49\linewidth,height=1.4cm]{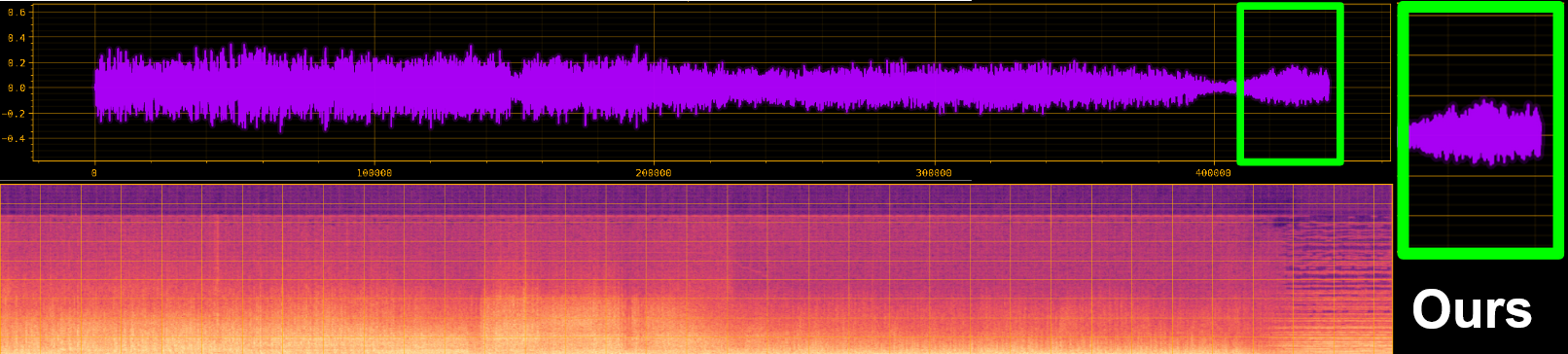}

\includegraphics[width=0.49\linewidth,height=1.4cm]{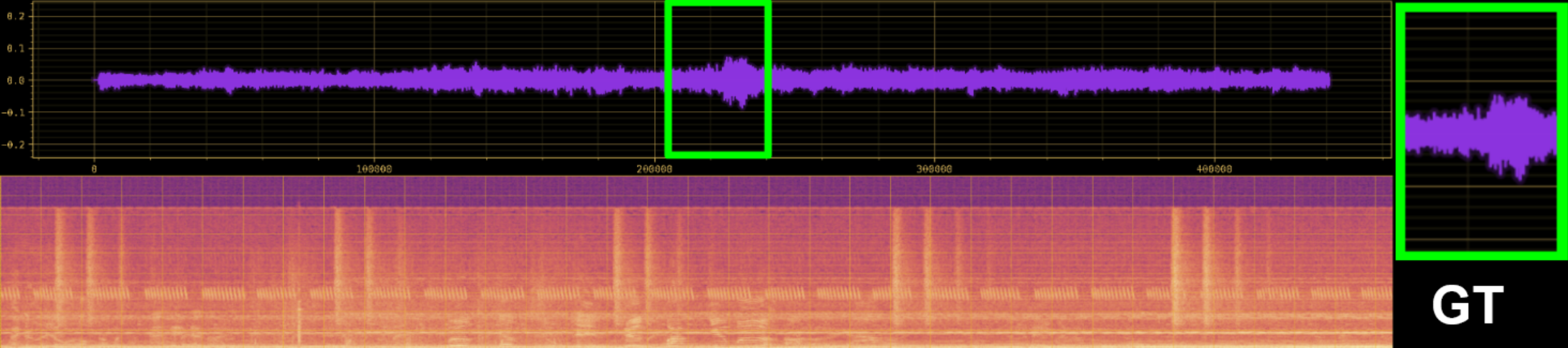}
\includegraphics[width=0.49\linewidth,height=1.4cm]{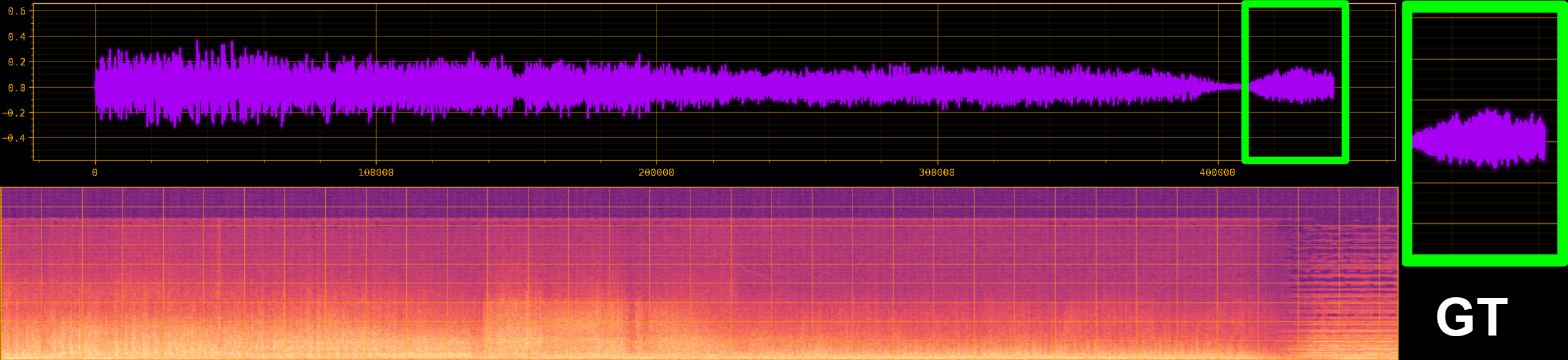}
\caption{Qualitative comparison Ours (VisAH-FM) vs VisAH. We achieve better balance by appropriately attenuating overly loud sources (SFX Column 1, Music Column 2).}
\label{fig:qualiFMVisah}
\end{figure}
% \vspace{-1em}
\iffalse
\begin{figure*}[!t]
    \centering
    \begin{subfigure}{0.48\textwidth}
        \centering
        \includegraphics[width=\linewidth]{plot_20.pdf}
    \end{subfigure}
    \hspace{0.02\textwidth}
    \begin{subfigure}{0.48\textwidth}
        \centering
        \includegraphics[width=\linewidth]{plot_7.pdf}
    \end{subfigure}

    \vspace{0.3em}

    \begin{subfigure}{0.48\textwidth}
        \centering
        \includegraphics[width=\linewidth]{plot_4.png}
    \end{subfigure}
    \hspace{0.02\textwidth}
    \begin{subfigure}{0.48\textwidth}
        \centering
        \includegraphics[width=\linewidth]{plot_6.pdf}
    \end{subfigure}

    \caption{Qualitative comparison of FM against Visah. Main differences are framed in red, and waveforms are overlapped for readability.}
    \label{fig:qualiFMVisah}
\end{figure*}
\fi

\paragraph{Subjective Test.} We conduct a subjective test to compare VisAH and VisAH-FM. Specifically, we ask participants to ``choose the audio that best aligns with the visual scene in terms of loudness balance and overall quality''. Thirteen participants evaluated the models on a set of nine Muddy-Mix and MovieGen \cite{moviegen} generated videos. VisAH-FM achieves a $45.55\pm9.46\%$ win rate versus $17.17\pm6.17\%$ for VisAH, with the remaining $37.28\%$ of comparisons judged as draws. These results underscore our method's effectiveness.
\section{Related work}
\label{sec:formatting}

\paragraph{Audio Remixing.}
Highlighting a mixed audio signal can be viewed as rebalancing its constituent sources—effectively translating one mixing style into another. Prior work in music production has examined this extensively~\cite{koo2023music,martinez2020deep,ramirez2019modeling,vanka2024diff}, exploring both traditional and creative strategies for shaping a track’s emotional and acoustic character. Reproducing a target mix often requires controlling source energy and applying audio effects to achieve stylistic consistency, whether through expert-driven pipelines~\cite{schlecht2022physical} or modern learning-based approaches~\cite{martinez2020deep,steinmetz2022style}. While most existing efforts emphasize music and instrument stems, these techniques do not generalize naturally to the diverse and highly dynamic sound compositions in real-world media. In contrast, VisAH~\cite{visah} extends the mixing paradigm beyond music to speech and cinematic sound effects, and leverages visual cues from video to guide source highlighting. While effective, this method relies on a discriminative mapping from input to output mixes, implicitly assuming a one-to-one correspondence between degraded audio mixture and professionally mixed audio—an assumption that may not hold in ambiguous and diverse audio contexts like movies. Concurrently, SemMix~\cite{semmix} studies LVLM-derived semantic prompts for visually-guided acoustic highlighting, which is complementary to our focus on flow matching.
\iffalse
\paragraph{Generative Modeling for Paired Translation}
Paired translation aims to learn a mapping between two domains when aligned pairs are available, enabling faithful translation while preserving content structure. Early approaches relied on supervised pixel-level reconstruction losses and encoder–decoder architectures to enforce cross-domain correspondence~\cite{isola2017image,zhu2017toward}. 
More recently, Latent Bridge Matching~\cite{LBM} proposes a principled alternative by learning a latent bridge between paired images space, where alignment is enforced via a flow matching objective rather than direct pixel regression. Built on top of a pretrained Stable Diffusion model, this framework benefits from strong visual priors, yielding high-fidelity and semantically consistent translations. 
%In addition, Latent Bridge Matching demonstrates conditional translation with an auxiliary modality (e.g., light position), highlighting its flexibility in leveraging external signals for controlled image transformations.
\fi
%\vspace{-1em}
\paragraph{Visually-conditioned Audio Generation.}
With the rapid progress in audio synthesis, several works have explored conditioning audio generation on video frames~\cite{moviegen,MMAudio,AudioX}. This setting is challenging due to the tight temporal coupling between visual events and sounds, as well as the high diversity of real-world audio scenes.
Current methods typically specialize in either speech or non-speech sounds (e.g., music, sound effects), limiting their generality. Consequently, these models are not directly suited for acoustic highlighting, which requires jointly modeling speech, music, and effects.
%\vspace{-1em}
\paragraph{Audio-Visual Source Separation.}
Early works in audio-visual source separation~\cite{soundPixel,lookListen,visualVoice} demonstrated that visual cues provide powerful guidance for isolating sound sources in videos.
More recent studies~\cite{davis,davisFlow,speechDiffusion} tackle this task using generative modeling, such as diffusion-based and flow-based frameworks, benefiting from their strong ability to model complex multimodal distributions and generate coherent audio conditioned on visual context.
Our work builds upon these trends by applying conditional flow matching to the audio remixing problem, focusing specifically on maintaining trajectory stability via a rollout loss and enabling early cross-modal fusion through adapter layers, which distinguishes our approach from prior diffusion-based conditioning pipelines.
\section{Conclusion}
We showed that generative modeling offers a more effective solution than discriminative approaches for visually informed acoustic highlighting. With a rollout loss and cross-modal conditioning module, our flow matching model outperforms prior works on the Muddy Mix dataset.

%We demonstrated that generative modeling provides a more effective solution than discriminative approaches for visually informed acoustic highlighting. By introducing a rollout loss and a cross-modal conditioning module, our flow-matching model significantly outperforms prior work on the Muddy-Mix dataset and exhibits improved stability and self-correction across inference steps.
%\\
%Despite these gains, our method incurs a higher computational cost due to multi-step inference and inherits limitations from CLIP/CLAP representations, which can cause failures when visual or audio cues are weak or semantically misaligned. Moreover, as Muddy-Mix is currently the only dataset available for this task, it would be valuable to assess the model on real-world data once such a dataset becomes available.

\newpage
\bibliographystyle{splncs04}
\bibliography{main}

\clearpage
\clearpage
\appendix
\section*{Supplementary Material}
\addcontentsline{toc}{section}{Supplementary Material}
\iffalse
\section{Demo Page: \textcolor{red}{\textbf{demo.html}}} 
The supplementary zip contains \texttt{demo.html} file (in the demo folder) showcasing the output of our VisAH-FM on example videos. \textbf{We strongly encourage readers to
visit this webpage - best experienced with headphones.} We recommend using Google Chrome. The demo is organized as follows:
\begin{itemize}
    \item We show examples from the Muddy Mix Dataset, where we compare the poorly balanced input, output of the discriminative baseline VisAH, proposed VisAH-FM and the original movie clip.
    \item Additionally, we demonstrate our method's utility as a post-processing stage for Video-to-Audio generation pipelines (e.g., MovieGen \cite{moviegen}), which may neglect subtle loudness imbalances across acoustic sources. Comparisons against VisAH are provided in this setting as well.
\end{itemize}
\fi 

\section{Additional Experiments}
\label{sec:addiExpe}

\subsection{Ablation of the Time Conditioning}
%When the timestep is fed to the model by the latent transformer, the encoder part of the model cannot access it, hence it might result in suboptimal results. Table \ref{tab:timestep} compares the performance of a model trained with time conditioning only in the latent transformer and another one where the timestep is fed through all the layers of the encoder (by summing the output of the layer with the encoding) in addition to the latent space conditioning.\\
%Interestingly, feeding the timestep through all the encoder layers does not improve the performance. We hypothesize that this is due to the fact that the encoder only extracts generic features about the audio, and the edition part is done in the latent transformer and decoder; hence, the model does make use of the timestep in the encoding part.
When the timestep is fed only to the latent transformer, the encoder does not have access to temporal information, which could theoretically lead to suboptimal representations. Table~\ref{tab:timestep} compares a model using timestep conditioning exclusively within the latent transformer against a model where the timestep is also injected into each layer of the encoder.
\\
The results indicate that adding timestep information to the encoder does not provide any measurable benefit. We hypothesize that the encoder mainly extracts general audio representations, while the actual editing is handled in the latent transformer and decoder. Therefore, the timestep information appears to be unnecessary within the encoder.
\begin{table}[!h]
    \caption{Ablation study on timestep encoding. Incorporating the timestep within the encoder does not improve performance compared with injecting it directly into the latent space.}
    \centering
    \scriptsize
    \setlength{\tabcolsep}{5pt}
    \begin{tabular}{lcccccc}
        \toprule
         Model & IB Score $\uparrow$& KLD $\downarrow$& LDif $\downarrow$ & Mag $\downarrow$& Env $\downarrow$& Was $\downarrow$\\
         \toprule
        Latent Transformer& \textbf{29.09} & \textbf{9.79} & 7.87 & 8.34 & 2.79 &0.65\\
        Encoder + Latent Transformer&29.02 &9.87&\textbf{7.83} &8.24&2.70&0.60 \\
         \bottomrule
    \end{tabular}
    
    \label{tab:timestep}
\end{table}

\subsection{No Warm Start}
%Table \ref{tab:noWarm} shows the performance of model trained with and without warm start (i.e. incorporating the source sample $x0$ into the model output to define the estimated vector field, as explained in Section \ref{setting}).\\
%The warm up proves always beneficial, as it allows for better utilization of the pretrained model. Interestingly, performance drops substantially when no rollout loss is applied, showing the inherent complexity of the flow matching setup for that particular task, without proper reliance on a pretrained model nor explicit rollout loss.
Table~\ref{tab:noWarm} reports the performance of models trained with and without warm start, \textit{i.e.}, incorporating the source sample $x_0$ into the estimated vector field as described in Section~\ref{sec:exp}. Warm start consistently proves beneficial by allowing better exploitation of the pretrained model. Performance degrades significantly when training with flow matching alone and without warm start, highlighting the difficulty of flow matching for this task without pretrained guidance or explicit rollout regularization.
\begin{table}[!h]
    \caption{Warm start ablation. Incorporating $x_0$ in the prediction significantly improves the performance, particularly when no rollout loss is applied.}
    \centering
    \scriptsize
    \setlength{\tabcolsep}{5pt}
    \begin{tabular}{lccccccc}
    \toprule
         Objective & Warm start & IB Score $\uparrow$& KLD $\downarrow$& LDif $\downarrow$ & Mag $\downarrow$& Env $\downarrow$& Was $\downarrow$\\
         \toprule
        FM + Rollout& \checkmark &\textbf{29.09} &\textbf{9.79}  & \textbf{7.87} & \textbf{8.46} & \textbf{2.79} & \textbf{0.64}\\
         FM only& \checkmark & 28.92 & 10.99 & 9.48 &  11.57&4.68 & 1.08\\
         \midrule
         FM + Rollout& \ding{55}& 29.28 & 10.65 & 8.41 & 8.71 & 2.80 & \textbf{0.64}\\
        FM only& \ding{55}& 28.75 & 16.52 & 12.73 & 15.54 & 4.56 &1.37 \\
         \bottomrule
    \end{tabular}
    
    \label{tab:noWarm}
\end{table}

\subsection{Impact of Number of Inference Steps for Rollout-only Model}
Table~\ref{tab:stepRoll} evaluates the model trained only using rollout loss for different number of inference steps. Consistent with the behavior observed in Figure~\ref{fig:cosinePlot}, the first step tends to align well with the ground-truth linear trajectory, leading to good performance when using a single inference step. Performance decreases as the number of steps increases due to trajectory non-linearity (because it forces interpolation to bigger step), except when using four steps — matching the training setup — which yields the best overall metrics.
\begin{table}[h]
    \caption{Ablation of the number of inference steps using the model trained with rollout loss only. Intermediate steps degrades performance, as the model learns non linear trajectory, interpolating from middle steps give noisy result.}
    \centering
    \scriptsize
    \setlength{\tabcolsep}{5pt}
    \begin{tabular}{ccccccc}
    \toprule
        Number of steps &IB Score $\uparrow$& KLD $\downarrow$& LDif $\downarrow$ & Mag $\downarrow$& Env $\downarrow$& Was $\downarrow$\\
        \midrule
         1& \textbf{29.11} & 10.09 & 7.73 & 8.17 & 2.84 & 0.66\\
         2& 28.86 & 10.17 & 8.36 & 9.15 & 2.92 & 0.70\\
         3& 28.78 & 10.68 & 9.42 & 10.34 & 0.32 & 0.84\\
        4& 28.94 & \textbf{9.92} & \textbf{7.71} & \textbf{8.13} & \textbf{2.78} & \textbf{0.62} \\
        \bottomrule
    \end{tabular}
    \label{tab:stepRoll}
\end{table}

\subsection{FAD}
Table~\ref{tab:fad} reports the FAD score for the baseline, the input (poorly mixed audio), and VisAH-FM. VisAH-FM performs twice as well as VisAH in FAD, which can be explained by the well-known quality of Flow Matching models to generate very faithful output, unlike discriminative models.

\begin{table}[!htbp]
    \caption{FAD scores}
    \centering
    \scriptsize
    \setlength{\tabcolsep}{5pt}
    \begin{tabular}{cc}
        \toprule
        Method & FAD$\downarrow$ \\
        \midrule
        Input & 21.31 \\
        VisAH & 3.61 \\
        VisAH-FM & \textbf{1.56} \\
        \bottomrule
    \end{tabular}
    \label{tab:fad}
\end{table}

\subsection{Statistical Significance}
We evaluate whether the improvement over VisAH is statistically significant on the Muddy Mix test set. Table~\ref{tab:stat_sig} reports p-values from paired Wilcoxon tests on per-example errors. All p-values are far below conventional significance thresholds, showing that the improvements are statistically significant across both signal-level and semantic metrics.

\begin{table}[h]
    \caption{Statistical significance of VisAH-FM improvements over VisAH on the Muddy Mix test set. We report p-values from paired Wilcoxon signed-rank tests.}
    \centering
    \scriptsize
    \setlength{\tabcolsep}{8pt}
    \begin{tabular}{cccc}
        \toprule
        Mag & LDif & IB Score & KLD \\
        \midrule
        $2.1\times10^{-78}$ & $1.7\times10^{-49}$ & $1.6\times10^{-7}$ & $1.6\times10^{-29}$ \\
        \bottomrule
    \end{tabular}
    \label{tab:stat_sig}
\end{table}
\subsection{Memory Usage}
The peak inference VRAM is unchanged relative to the corresponding non-rollout model because the compared models use the same number of parameters. The additional memory cost mainly appears during training, where the rollout loss requires backpropagating through multiple discretization steps. For one training iteration with batch size $3$ and four discretization steps, peak VRAM increases from $20.59$~GB without rollout to $43.63$~GB with rollout. This reflects the computational cost of the rollout regularization, while leaving the inference memory footprint unchanged.

\section{Signal Metrics for All Ablations}
Table~\ref{tab:conditioningMetric} and \ref{tab:comparison_rollout_fullMet} report all metrics for the conditioning and rollout ablations, respectively. Most signal-level metrics follow trends similar to those of the semantic metrics, except for VisAH-FM trained with text-only inputs, which outperforms the other methods on signal metrics while underperforming on semantic metrics. Table~\ref{tab:HP_supp} further reports the full signal-level hyperparameter sensitivity, confirming that performance remains stable across rollout weights and horizons, with only a small degradation when increasing the number of inference steps.
\iffalse
\begin{table}[h]
    \caption{Ablation on the conditioning}

    \centering
    \begin{tabular}{ccccccc}
        \toprule
        Model & IB Score & KLD & LDif & Mag & Env & Was\\
        \midrule
        FM CMT Audio+Text & 29.10 & 9.73 & 7.90 & 8.25 &2.74 & 0.63\\
        FM CMT Audio & \textbf{29.12} & \textbf{9.70} & \textbf{7.77} & 8.28 &2.74 & 0.63\\
        FM CMT Text & 29.08 & 9.76 &7.85  & 8.34 & 2.74& 0.63\\
        FM Vision & 29.09 & 9.79 & 7.87 & 8.46 & 2.79 & 0.64\\
        FM Text & 28.88 & 10.50 & 8.13 & \textbf{8.22} & \textbf{2.65} & \textbf{0.54}\\
        \bottomrule
    \end{tabular}
    \label{tab:conditioning_full}
\end{table}
\fi

\begin{table}[!t]
\caption{Ablation on the conditioning. We compare different conditioning methods (with and without adapter) and different modalities: audio (A), vision (V), and text (T).}
    \centering
    \scriptsize
    \setlength{\tabcolsep}{5pt}
    \begin{tabular}{cccccccc}
        \toprule
        Adapter & Modality & IB Score $\uparrow$& KLD $\downarrow$& LDif $\downarrow$ & Mag $\downarrow$& Env $\downarrow$& Was $\downarrow$\\
        \midrule
        \checkmark & V+A& \textbf{29.12} & \textbf{9.70} & \textbf{7.77} & 8.28 &2.74 & 0.63\\
        \checkmark & V + T & 29.08 & 9.76 &7.85 & 8.34 & 2.74& 0.63\\
        \checkmark & V+T+A &29.10 & 9.73 & 7.90&8.25 &2.74 & 0.63\\
        \ding{55} & V & 29.09 & 9.79 & 7.87  & 8.46 & 2.79 & 0.64\\
        \ding{55} & T &28.88 & 10.50 & 8.13 & \textbf{8.22} & \textbf{2.65} & \textbf{0.54}\\
        \bottomrule
    \end{tabular}

    \label{tab:conditioningMetric}
\end{table}

\begin{table}[h]
    \caption{Ablation of the rollout component. We compare different
standard methods used to stabilize predictions with our rollout proposal. All models use standard CLIP conditioning (w/o adapter).}
    \centering
    \scriptsize
    \setlength{\tabcolsep}{5pt}
    \begin{tabular}{lcccccc}
    \toprule
         Model & IB Score $\uparrow$& KLD $\downarrow$& LDif $\downarrow$ & Mag $\downarrow$& Env $\downarrow$& Was $\downarrow$\\
         \toprule
         FM + Rollout& \textbf{29.09} &\textbf{9.79}  & 7.87 & 8.46 & 2.79 & 0.64\\
         \midrule
         FM only& 28.92 & 10.99 & 9.48 &  11.57&4.68 & 1.08\\
         Rollout& 28.94 & 9.92 & \textbf{7.71} & \textbf{8.13} & \textbf{2.78} & \textbf{0.62}\\
         FM + Consistency & 28.40 & 14.58 & 10.30 & 13.54 & 3.98 & 1.15\\
         FM Weighted & 28.86 & 11.16 & 9.44 & 11.83 & 3.68& 0.99\\
         Bridge Matching & 29.04 & 10.85 & 9.62 & 10.98 & 3.51& 0.91\\
         %FM with perceptual loss &29.63 & 8.89  & 8.17 & 10.46 & 3.30 & 0.84\\
         %RM Foll03 percept & 29.78 & 8.78 & 8.04 & 10.69 & 3.35 & 0.84\\
         \bottomrule
    \end{tabular}
    \label{tab:comparison_rollout_fullMet}
\end{table}

\iffalse
\begin{table}[H]
    \centering
    \begin{tabular}{ccccccc}
         Model & IB Score & KLD & LDif & Mag & Env& Was\\
         \toprule
         rollout 0.3- 10 steps-H4& 29.10 & 9.79 & 8.35 & 8.58 & 2.81&0.66 \\
         rollout 0.1& 29.12 & 9.85 & 8.02 & 8.41 & 2.78 &0.64\\
         Rollout 0.3& 29.09 &9.79  & 7.87 & 8.46 & 2.79 & 0.64\\
         Rollout 0.5& 29.12 & 9.93 & 7.80 & 8.43 & 2.79& 0.64\\
         Rollout 1.0& 29.04 & 9.82 & 7.84 & 8.47 & 2.81& 0.65\\
         R0.1 horizon 2 & 29.08 & 9.68 & 7.94 &8.45  & 2.76 & 0.63\\
         No Warm start W roll& 29.28 & 10.65 & 8.41 & 8.71 & 2.80 & 0.64\\
        No Warm start W/o roll& 28.75 & 16.52 & 12.73 & 15.54 & 4.56 &1.37 \\
        rollout03 time enc& 28.99 & 9.80 & 7.87 & 8.34 & 2.79 &0.65\\
        rollout03CMT time enc&29.02 &9.87&7.83 &8.24&2.70&0.60
         \bottomrule
    \end{tabular}
    
    \caption{Hyperparameter sensitivity}
    \label{tab:HP_full}
\end{table}
\fi

\begin{table}[h]
    \caption{Hyperparameter sensitivity. We vary the impact of the rollout loss as well as its horizon and the number of total steps used for inference.}
    \centering
    \scriptsize
    \setlength{\tabcolsep}{5pt}
    \begin{tabular}{ccccccccc}
    \toprule
         $\lambda$ & Horizon  & Steps & IB Score $\uparrow$ & KLD $\downarrow$ & LDif $\downarrow$ & Mag $\downarrow$ & Env $\downarrow$ & Was $\downarrow$\\
    \toprule
         0.1 & 4 & 4 & \textbf{29.12} & 9.85 & 8.02 & \textbf{8.41} & 2.78 & 0.64 \\
         0.3 & 4 & 4 & 29.09 & 9.79 & 7.87 & 8.46 & 2.79 & 0.64 \\
         0.5 & 4 & 4 & \textbf{29.12} & 9.93 & \textbf{7.80} & 8.43 & 2.79 & 0.64 \\
         1.0 & 4 & 4 & 29.04 & 9.82 & 7.84 & 8.47 & 2.81 & 0.65 \\
    \midrule
         0.1 & 2 & 4 & 29.08 & \textbf{9.68} & 7.94 & 8.45 & \textbf{2.76} & \textbf{0.63} \\
         0.3 & 4 & 10 & 29.10 & 9.79 & 8.35 & 8.58 & 2.81 & 0.66 \\
    \bottomrule
    \end{tabular}

    \label{tab:HP_supp}
\end{table}

\section{Linearity of Trajectories in Semantic Space}
The analysis in Section~\ref{sec:behave} showed that the rollout loss helps linearize the inference trajectory. However, when its contribution becomes too dominant, the trajectories exhibit non-linear behavior again. This analysis was conducted in the time–frequency domain. Here we analyze linearity of trajectories in a semantic representation space, where each inference step can affect how the audio is interpreted.

Table~\ref{tab:linearityPasst} reports the variance of the estimated vector fields as well as the average discrete curvature of the model outputs, measured in the PaSST embedding space~\cite{PASST}. The observations mirror those made in the time–frequency domain: adding the rollout loss alongside flow matching reduces trajectory curvature, while either increasing its weight or removing the flow matching loss altogether leads to more non-linear trajectories.

In a real-world scenario, such behavior is beneficial for models trained with both flow matching and rollout loss. It might allow a user to flexibly control the desired amount of remixing by stopping inference after a chosen number of steps, enabling semantic manipulation along a smooth editing path. 
\begin{table}[h]
    \caption{Linearity metrics in the PASST space. Adding the rollout loss linearized the inference trajectories, but applying it without the flow matching loss results in non-linear trajectories.}
    \centering
    \scriptsize
    \setlength{\tabcolsep}{5pt}
    \begin{tabular}{lcc}
    \toprule
        Training objective & $V(v_{\theta}(t))$ & Curvature \\
        \toprule
        FM & 0.0252 & 0.9188\\
        FM + RollOut ($\lambda=0.1$) & 0.0052 & 0.4390 \\
        FM + RollOut ($\lambda=0.3$)  & 0.0052 & 0.4397 \\
        RollOut & 0.0218 & 0.9406\\
        \bottomrule
    \end{tabular}
    \label{tab:linearityPasst}
\end{table}

\iffalse
\begin{table}[H]
    \centering
    \begin{tabular}{cccc}
    \toprule
        Training objective & $V(v_{\theta}(t))$ & Curvature & Tortuosity \\
        \toprule
        Regular FM & 0.0252 & 0.9188 & 2.8003\\
        RollOut 0.1 & 0.0052 & 0.4390 & 2.6913 \\
        RollOut 0.3 & 0.0052 & 0.4397 & 2.7316 \\
        RollOut Only & 0.0218 & 0.9406 &2.7153 \\
        \bottomrule
    \end{tabular}
    \caption{In pASST space}
    \label{tab:linearityPasst}
\end{table}

\begin{figure}[H]
    \centering
    \includegraphics[width=0.5\linewidth]{linearity.png}
    \caption{Linearity past space}
    \label{fig:enter-label}
\end{figure}
Linearity WAV

Regular
Vts 0.0001
curvatureL 0.6905
p_lenL 2.5498

Roll03
Vts 1.6605e-05
curvatureL
0.2461
P_len 2.5114

Rollout10
Vts 2.8140e-05
curvatureL 0.3255
P len 2.5356

Rollonly
Vts 0.0003
urvatureL 0.7047
P_len 2.5279
\fi
\section{Dataset details}
\label{sec:dataset}
We train and evaluate our models using the Muddy Mix dataset, constructed from movie clips that contain professionally mixed audio aligned with carefully edited visuals. The dataset leverages the observation that films inherently provide high-quality audio–visual synchronization, which serves as a form of free supervision for learning how audio should be highlighted relative to visual content. Specifically, the authors collect clips from the Condensed Movie Dataset (CMD) and generate training pairs through a pseudo-data pipeline that simulates poorly mixed audio. This process involves three steps: (1) separating the original high-quality movie audio into individual sources, (2) adjusting the levels of these sources to disturb the intended balance, and (3) remixing them to create a degraded input while keeping the original mix as the ground-truth target.
The remixing process consists of strongly decreasing the loudness of the loudest source, then randomly selecting another source whose loudness is increased, and finally slightly reducing the loudness of the remaining source (three sources are always considered: music, speech, and sound effects).
\\
The final dataset consists of 15,078/1,927/1,789 clips for train/validation/test sets, respectively. Note that, in our experiments, we remix the audios on-the-fly, artificially augmenting the diversity an creating multiple inputs for the same output.
\section{Acoustic source-specific analysis}
%In order to analyse the main sources of error of VisaAH-FM, we performed a two-way ANOVA, using the magnitude distance as the target variable. 
For each video in the test set, we compute the magnitude error and analyze its relationship with the boosted and heavily attenuated sources in the poorly mixed input. This enables us to test the main effects of enhancement and suppression, as well as their interaction, thereby assessing both component-wise and combination-specific errors of VisAH-FM. Table \ref{tab:sourcesError} reports the average errors, measured as the magnitude distance between the predicted and ground-truth audio, depending on which source is boosted or heavily attenuated. 
To facilitate interpretation, we ignore the slightly attenuated source (see Section \ref{sec:dataset} for poorly-mixed input creation details). 
While it does well in most scenarios, the model produces notably larger errors when music is boosted while speech is attenuated, suggesting that re-synthesizing speech masked by loud music is particularly challenging. In contrast, the model performs better when speech is enhanced and sound effects are attenuated. One possible explanation is that loud speech does not mask sound effects as strongly, allowing the model to more accurately restore the loudness of the sound effects.
\begin{table}
    \centering
    \caption{Magnitude distance between predicted and ground truth audio, depending on the sources boosted or (heavily) attenuated in the poorly balanced input.}
    \scriptsize
    \setlength{\tabcolsep}{5pt}
    \begin{tabular}{llc}
        \toprule
        Boosted & Attenuated & Mag $\downarrow$ \\
        \midrule
        Music & SFX & 0.07 \\
        Music & Speech & 0.14 \\
        SFX & Music & 0.06 \\
        SFX & Speech & 0.08 \\
        Speech & Music & 0.07 \\
        Speech & SFX & 0.04 \\
         \bottomrule
    \end{tabular}
    \label{tab:sourcesError}
\end{table}

\section{Experiments on the Fixed Muddy Mix Dataset}
\label{sec:nonSampled}
%In order to evaluate the impact of the on-the-fly sampling of the sources (i.e., choosing randomly which source to enhance for each audio at each iteration), we trained VisAH-FM model on the fixed, pre-saved Muddy Mix dataset.\\
%Table \ref{tab:fixedMudy} shows the results of our model trained on that particular dataset. While the overall performance decreases, the main conclusion remains the same: the flow matching outperforms the discriminative model, and the rollout loss proves useful.
To evaluate the impact of on-the-fly source sampling (i.e., randomly selecting which source to enhance for each audio example at each iteration), we trained the VisAH-FM model on the fixed, pre-generated Muddy Mix dataset, that was originally proposed in \cite{visah}.
Table~\ref{tab:fixedMudy} presents the results obtained on this variant of the dataset. The main observations remain consistent. As expected, the overall performance decreases relative to training with random sampling: flow matching outperforms the discriminative baseline, and the rollout loss provides additional improvements.
Interrestingly, our model (even trained sub-optimally on the Fixed Muddy-Mix dataset, with one-to-one mapping) still outperforms SemMix~\cite{semmix}, showing the effectiveness of the generative approach.

%(GT IB score 0.300769) on our resample ds
\begin{table}[h]
    \caption{Results on the fixed Muddy Mix dataset. Flow matching setup improves performance, and adding the rollout loss stabilizes inference, as also observed for on-the-fly dataset sampling.}
    \centering
    \scriptsize
    \setlength{\tabcolsep}{5pt}
    \begin{tabular}{lcccc}
    \toprule
        Model & IB Score $\uparrow$& KLD $\downarrow$& LDif $\downarrow$ & Env $\downarrow$\\
        \midrule
        VisAH Text & 29.00 & 11.02 & 9.23 & 0.035 \\
        \midrule
        FM & 29.27 & 10.91 & 8.84 & 0.034\\
        FM + Rollout ($\lambda=0.1$) & 29.15 & 10.60 & \textbf{8.82} & \textbf{0.032}\\
        FM + Rollout ($\lambda=0.3$) & \textbf{29.29} & \textbf{10.48} & 8.83 & \textbf{0.032}\\
        Rollout & 29.18 & 11.38 & 8.87 & 0.033\\
        \bottomrule
        SemMix \cite{semmix} & 28.93 & 10.95 & - & 0.034 \\
        \bottomrule
    \end{tabular}

    \label{tab:fixedMudy}
\end{table}

\section{Limitations and Future Work}
VisAH-FM delivers meaningful performance gains compared to the discriminative VisAH, but it is computationally more demanding and inherits limitations from the underlying CLIP/CLAP representations, which may lead to failures when audio–visual cues are weak. Future work should evaluate the model on real-world data once such datasets become available.
Additionally, our current VisAH-FM is trained on paired data, requiring artificial mixing of audio signals. Training with unpaired data would enable the model to directly leverage real-world inputs, significantly increasing its practical applicability. %We leave this direction for future work

\section{In-depth Qualitative Analysis}
\begin{figure}
    \centering
    \includegraphics[width=0.6\linewidth]{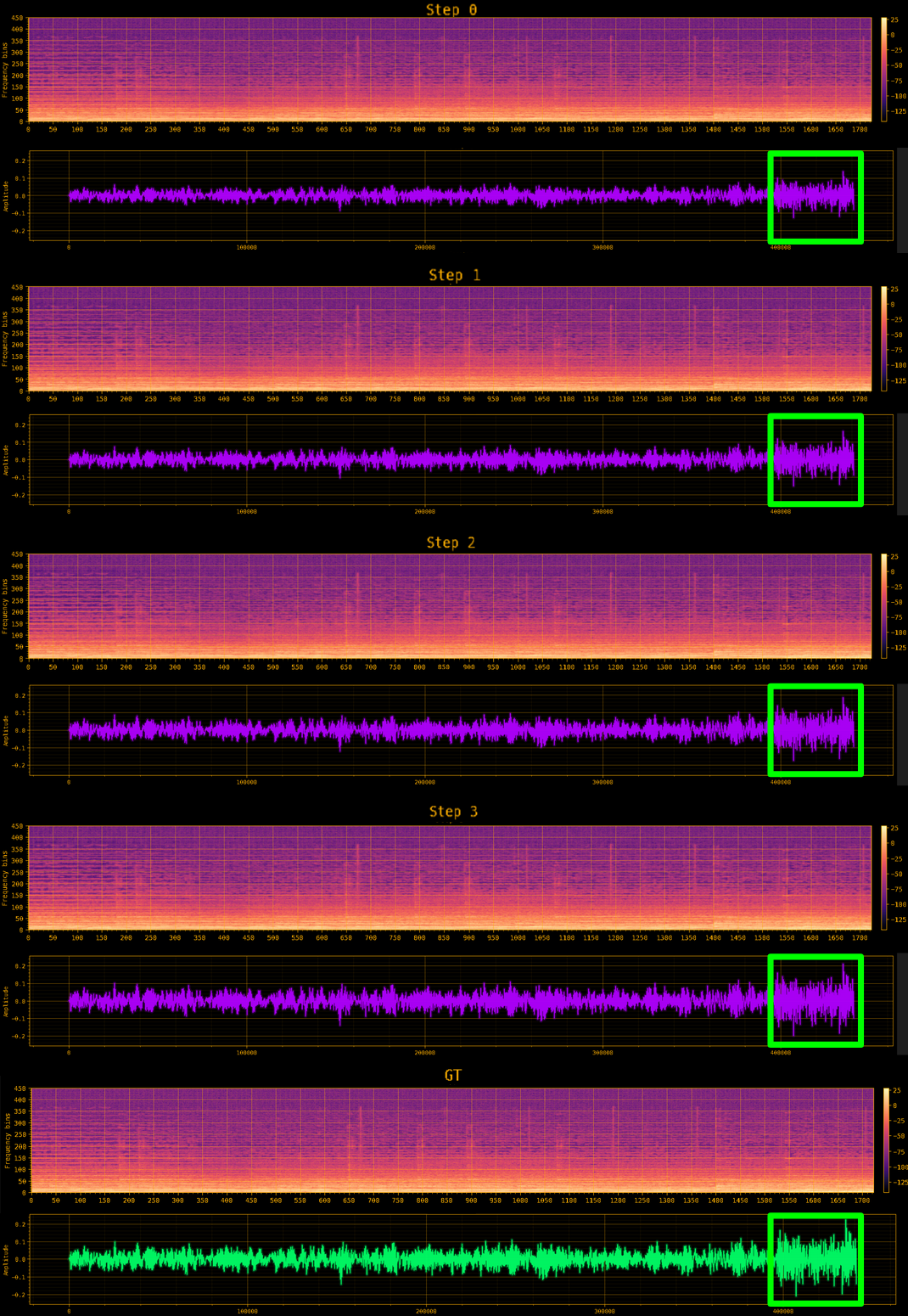}
        \caption{The output of VisAH-FM at different steps provide different level of remixing}
    \label{fig:stepPred}
\end{figure}

Figure \ref{fig:stepPred} shows the progressive remixing of the audio across the steps. The process evolves smoothly as the flow matching goes from poorly mixed audio to well mixed audios (as opposed to starting from noise. As such, the output of each step provides a realistic audio that corresponds to different degree of remixing.

\begin{figure}
    \centering
    \includegraphics[width=0.5\linewidth]{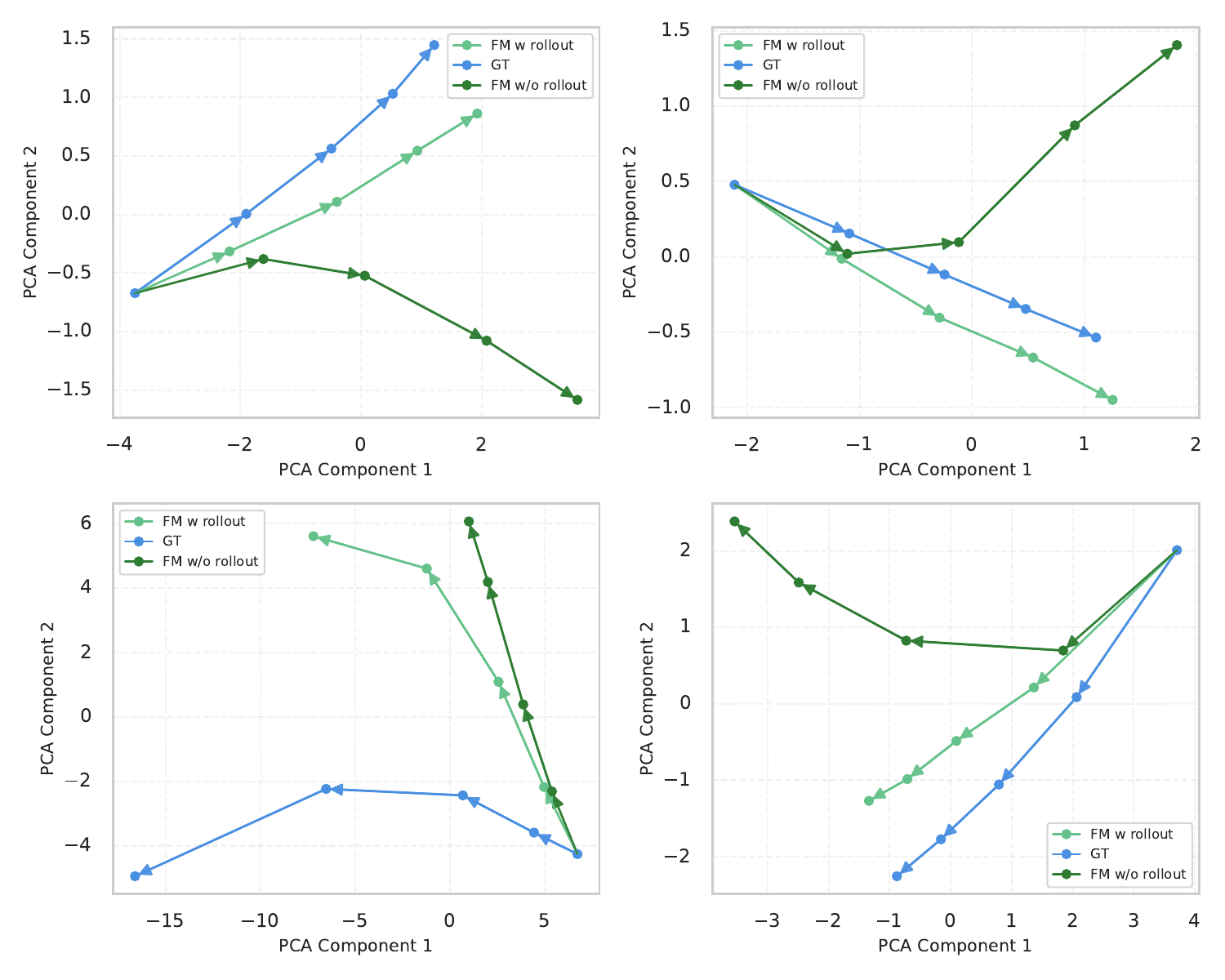}
    \caption{Trajectories with and without rollout loss, in the PASST space.}
    \label{fig:trajPasst}
\end{figure}
Figure~\ref{fig:trajPasst} visualizes PCA-projected trajectories in the PASST~\cite{PASST} embedding space for models trained with and without the rollout loss, alongside the ground-truth trajectory. Both models follow a similar direction at the first step, but clear differences emerge by the second step: the flow matching-only model quickly deviates and produces noisy trajectories, whereas the rollout-trained model remains close to the ground truth. Moreover, even when both models initially move in an incorrect direction, the rollout-trained model partially corrects its path over subsequent steps, showing robustness to early prediction errors.

\begin{figure*}[t]
    \centering
    
    % Row 1
    \begin{subfigure}{0.48\textwidth}
        \centering
        \includegraphics[width=\linewidth]{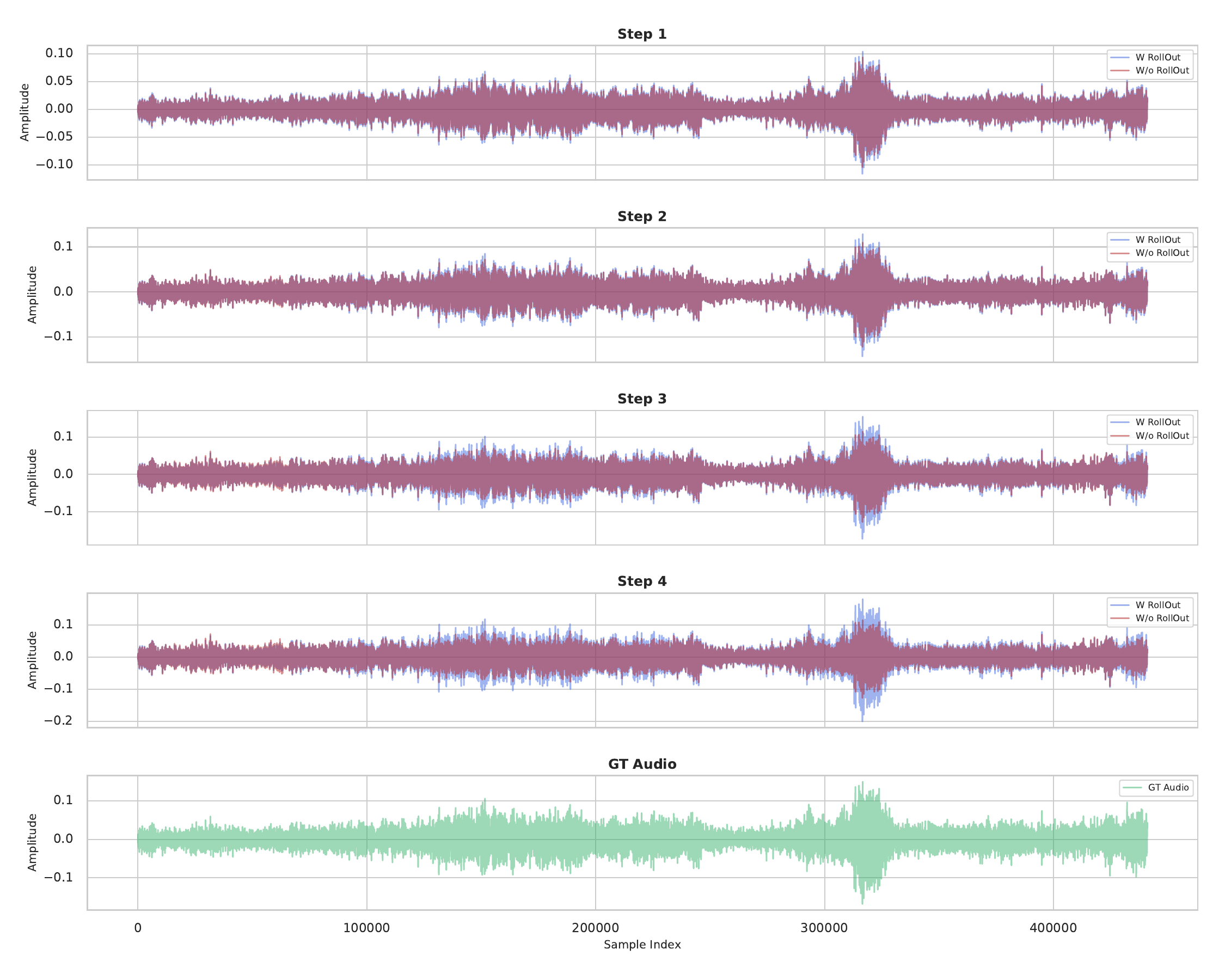}
        %\caption{Caption 1}
        \label{fig:rollout_sub1}
    \end{subfigure}
    \hfill
    \begin{subfigure}{0.48\textwidth}
        \centering
        \includegraphics[width=\linewidth]{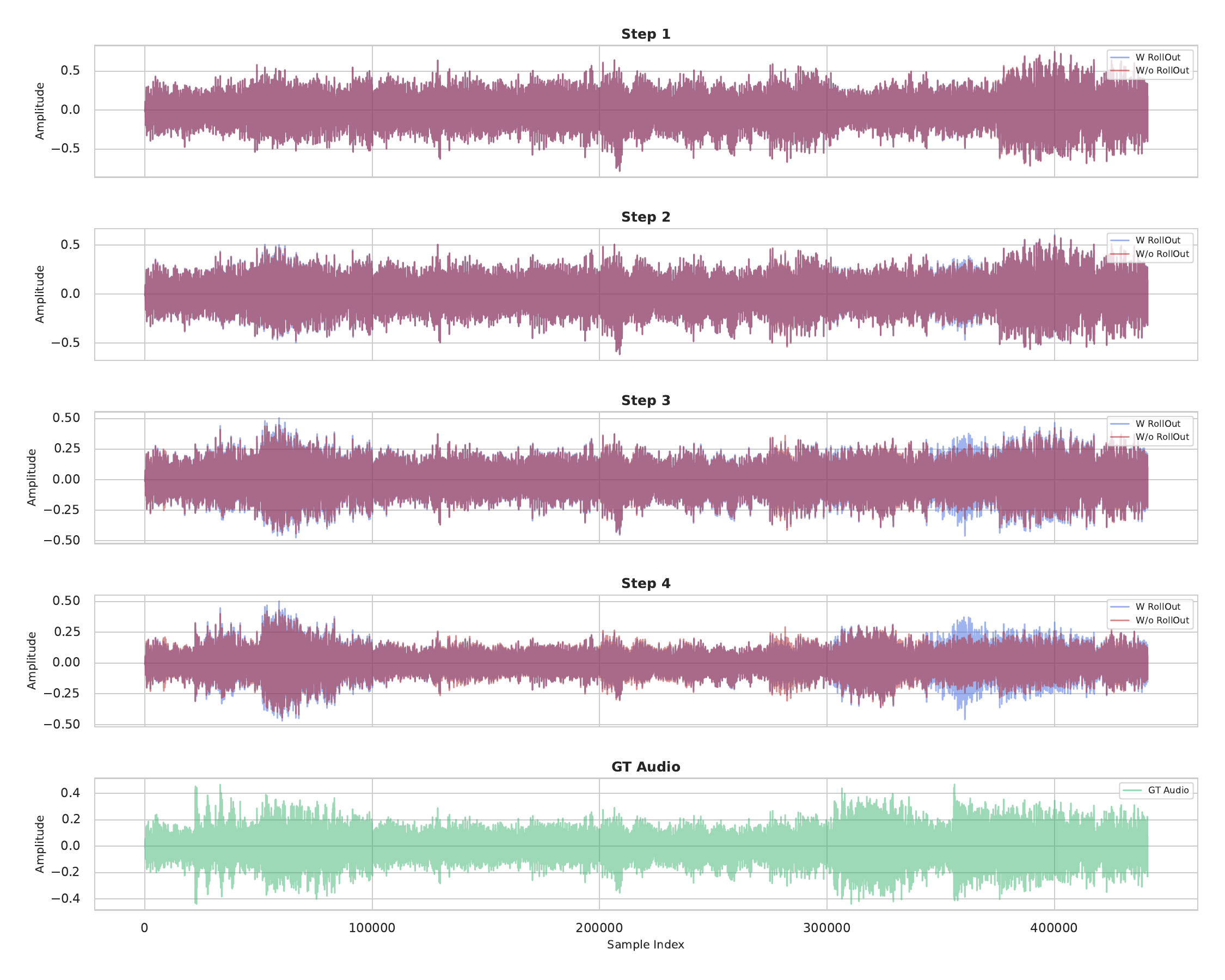}
        %\caption{Caption 2}
        \label{fig:rollout_sub2}
    \end{subfigure}

    \vspace{0.6em} % space between rows

    % Row 2
    \begin{subfigure}{0.48\textwidth}
        \centering
        \includegraphics[width=\linewidth]{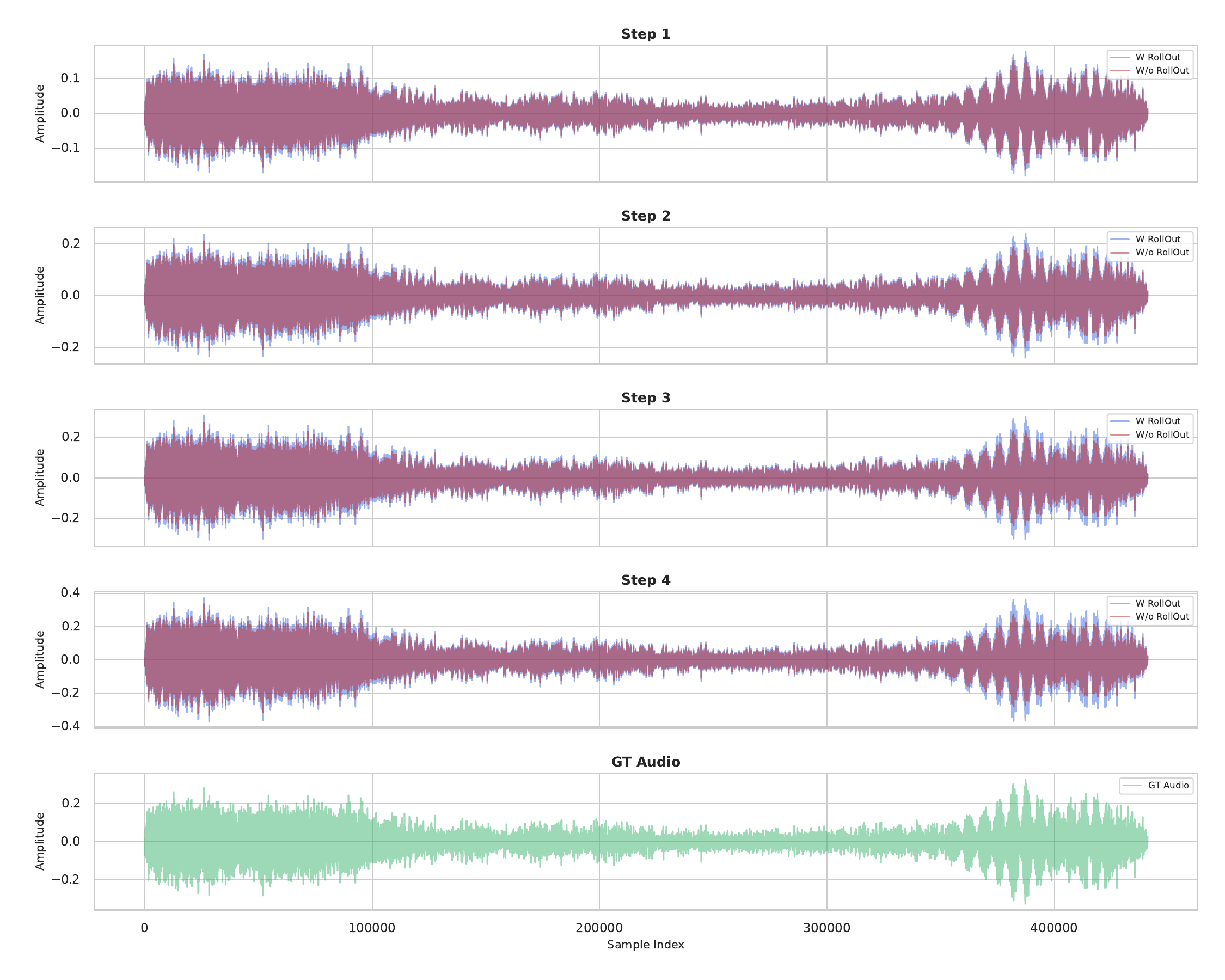}
        %\caption{Caption 3}
        \label{fig:rollout_sub3}
    \end{subfigure}
    \hfill
    \begin{subfigure}{0.48\textwidth}
        \centering
        \includegraphics[width=\linewidth]{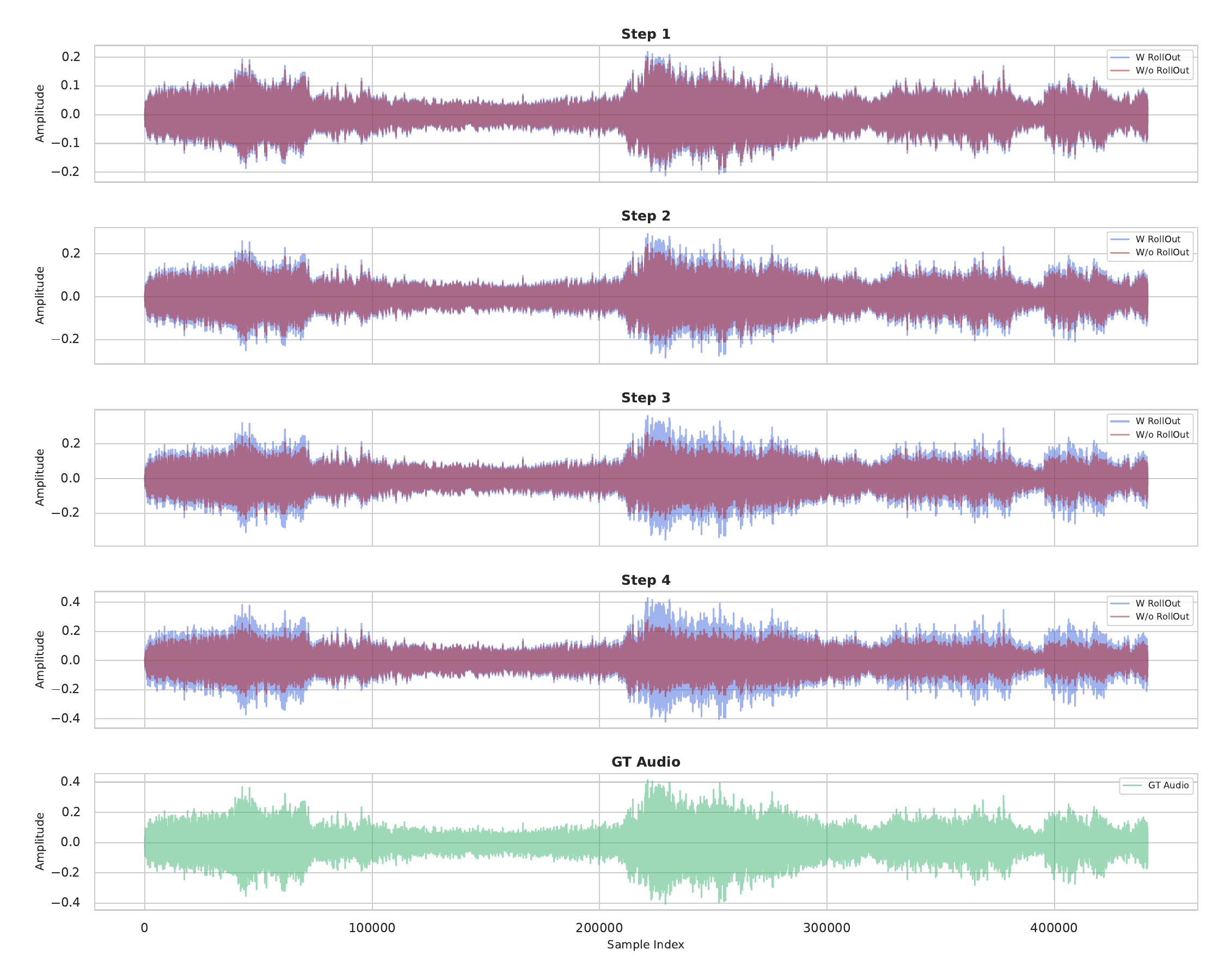}
        %\caption{Caption 4}
        \label{fig:rollout_sub4}
    \end{subfigure}

    \vspace{0.6em}

    % Row 3
    \begin{subfigure}{0.48\textwidth}
        \centering
        \includegraphics[width=\linewidth]{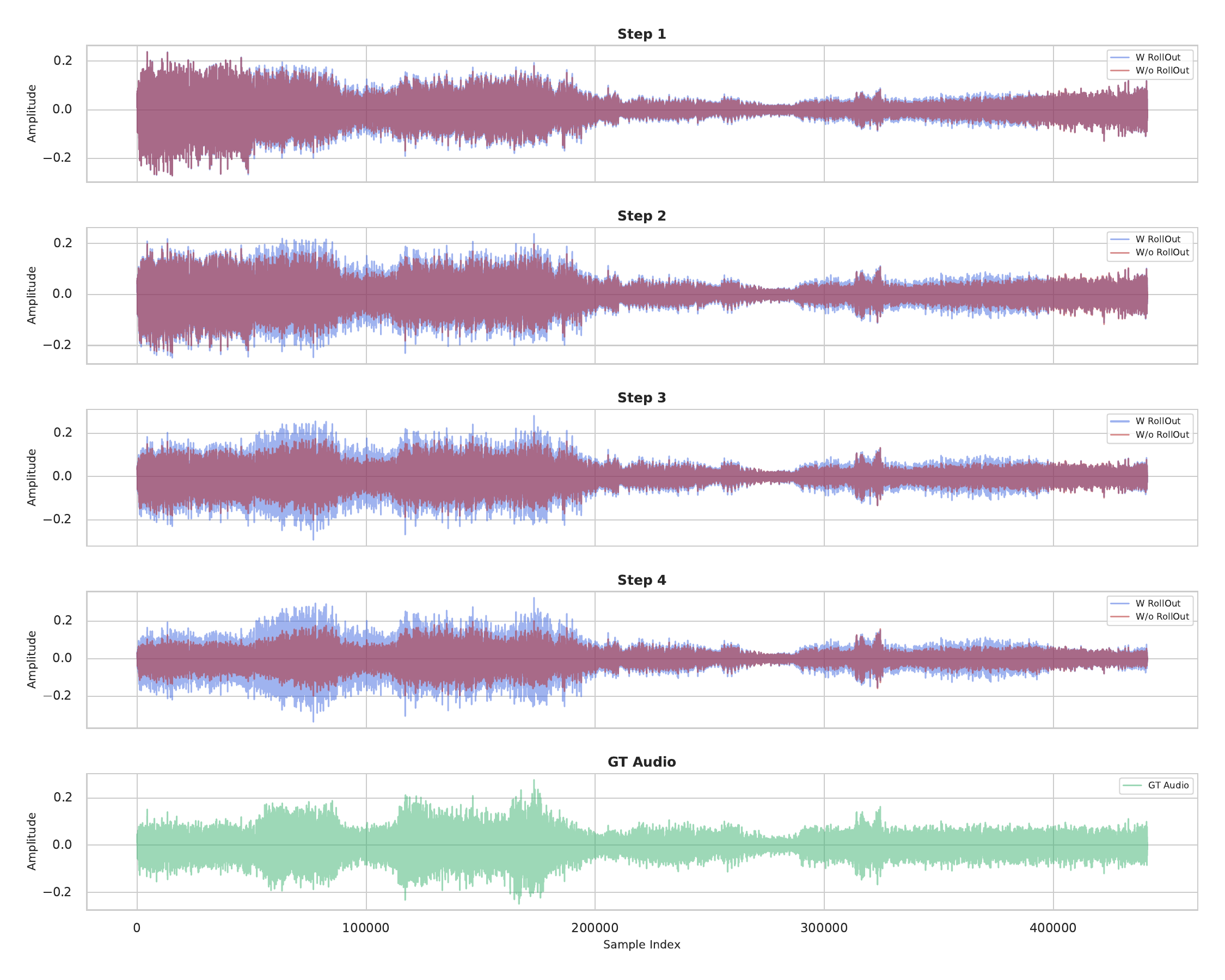}
        %\caption{Caption 5}
        \label{fig:rollout_sub5}
    \end{subfigure}
    \hfill
    \begin{subfigure}{0.48\textwidth}
        \centering
        \includegraphics[width=\linewidth]{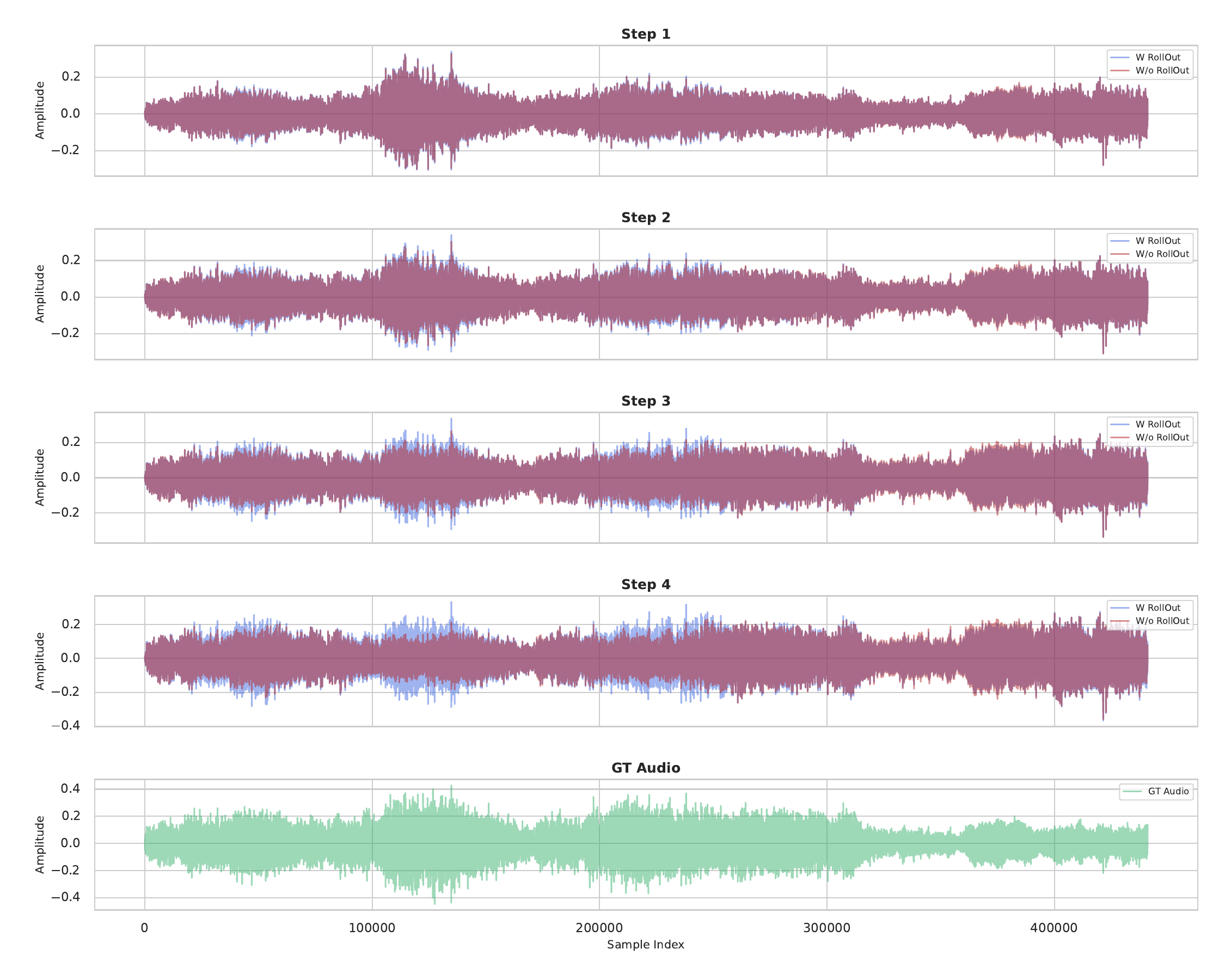}
        %\caption{Caption 6}
        \label{fig:rollout_sub6}
    \end{subfigure}

    \caption{Qualitative comparison of VisAH-FM models trained with and without rollout loss.}
    \label{fig:fullRoll}
\end{figure*}

\begin{figure*}[t]
    \centering
    
    % Row 1
    \begin{subfigure}{0.48\textwidth}
        \centering
        \includegraphics[width=\linewidth]{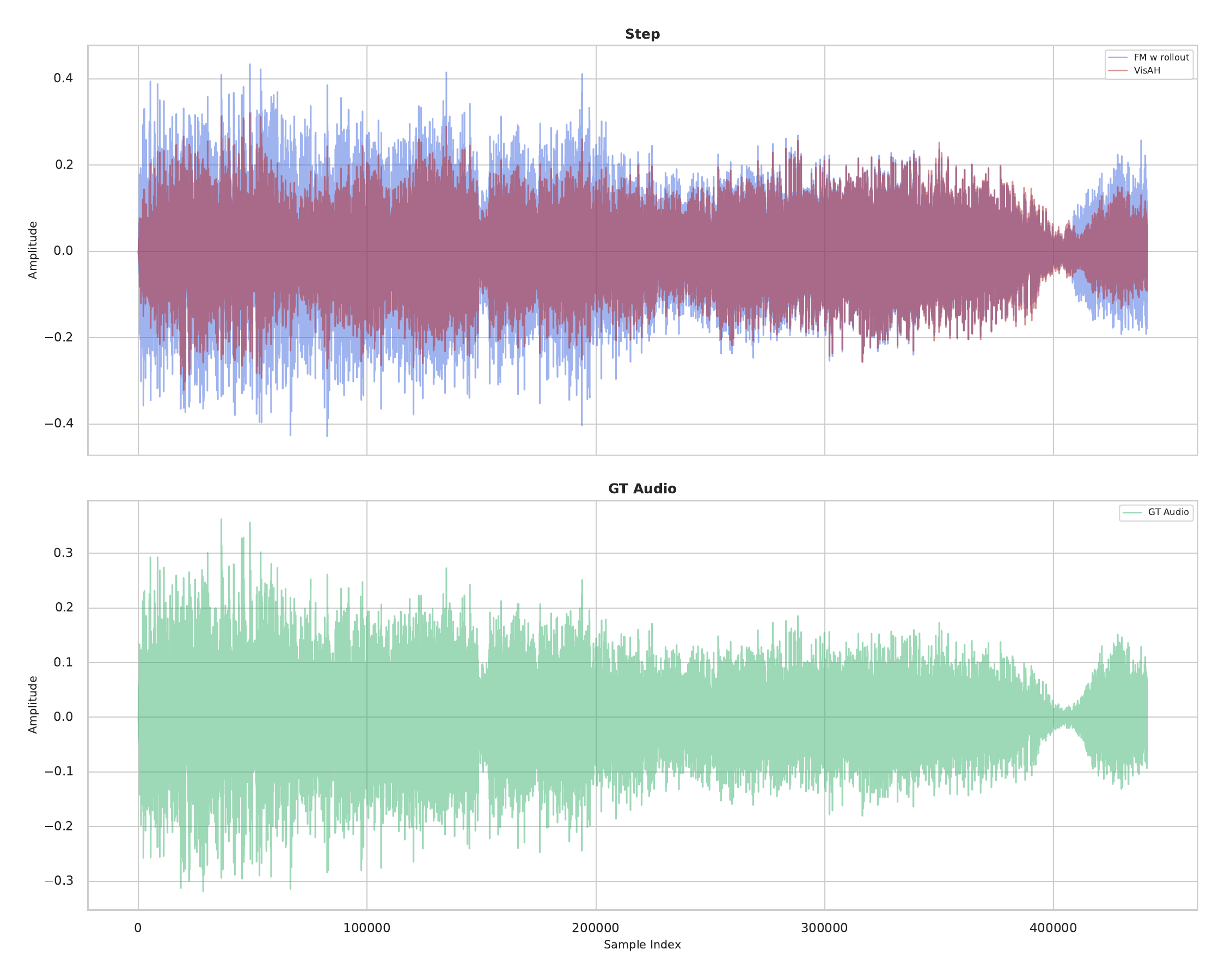}
        \label{fig:visah_sub1}
    \end{subfigure}
    \hfill
    \begin{subfigure}{0.48\textwidth}
        \centering
        \includegraphics[width=\linewidth]{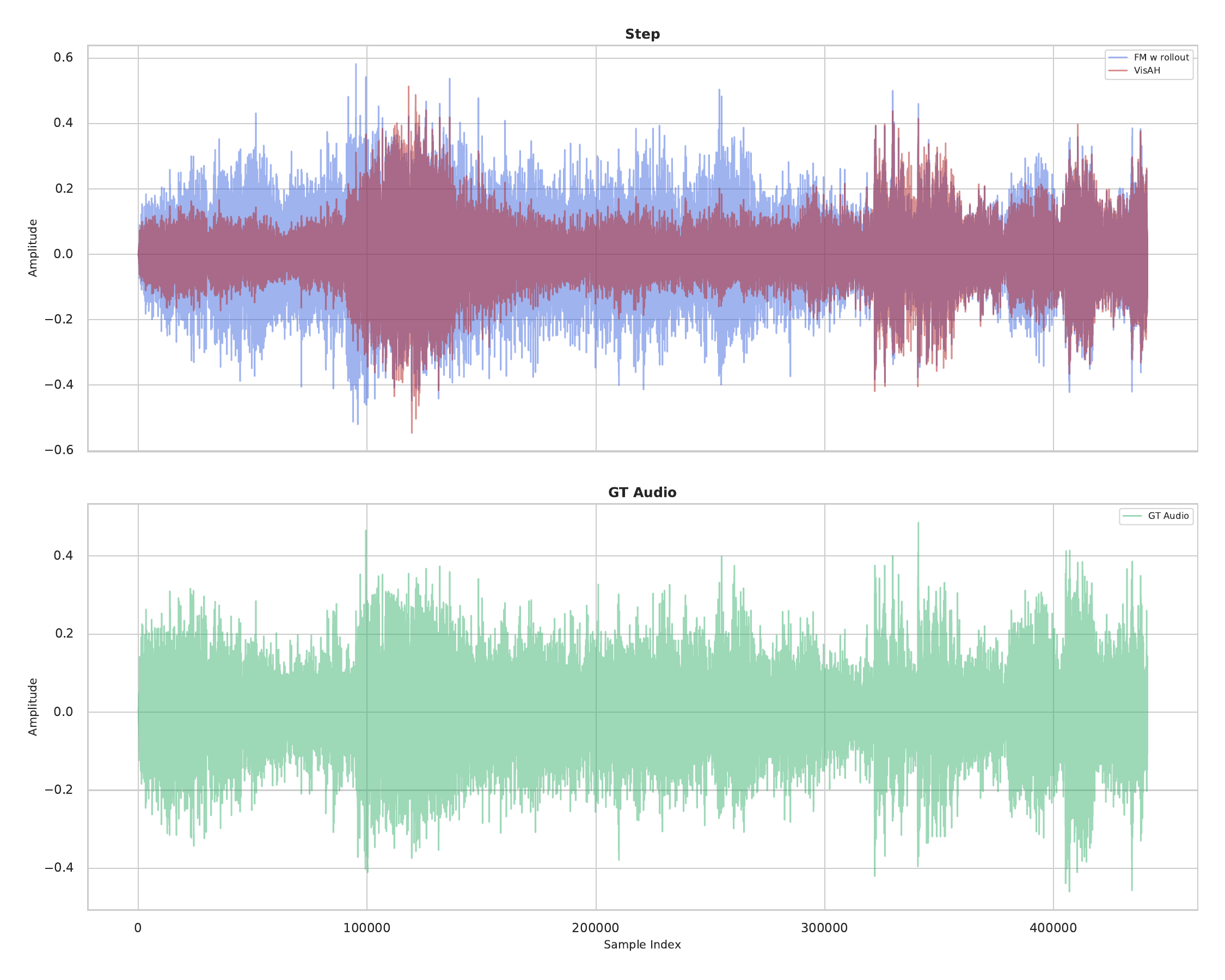}
        \label{fig:visah_sub2}
    \end{subfigure}

    \vspace{0.6em} % space between rows

    % Row 2
    \begin{subfigure}{0.48\textwidth}
        \centering
        \includegraphics[width=\linewidth]{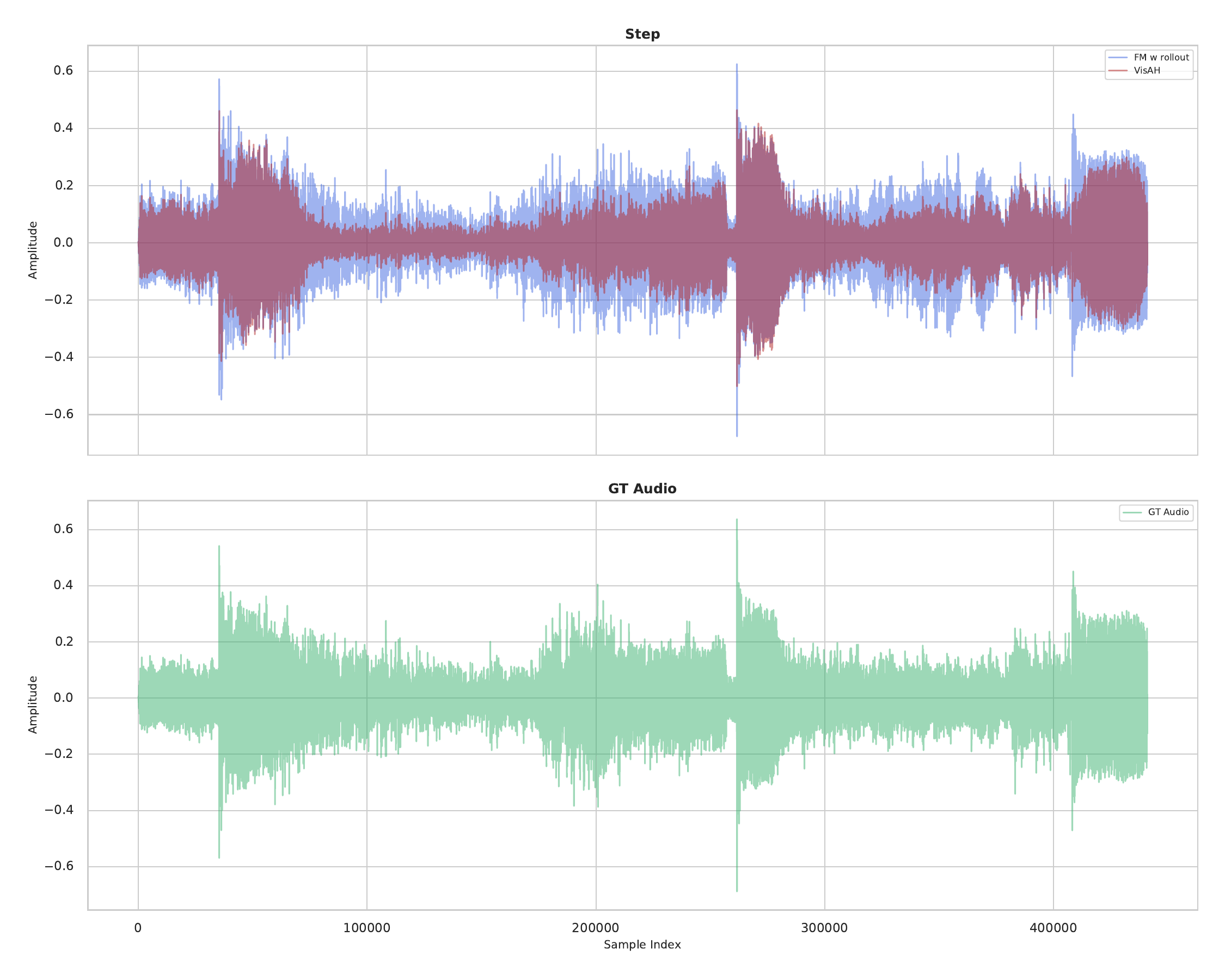}
        \label{fig:visah_sub3}
    \end{subfigure}
    \hfill
    \begin{subfigure}{0.48\textwidth}
        \centering
        \includegraphics[width=\linewidth]{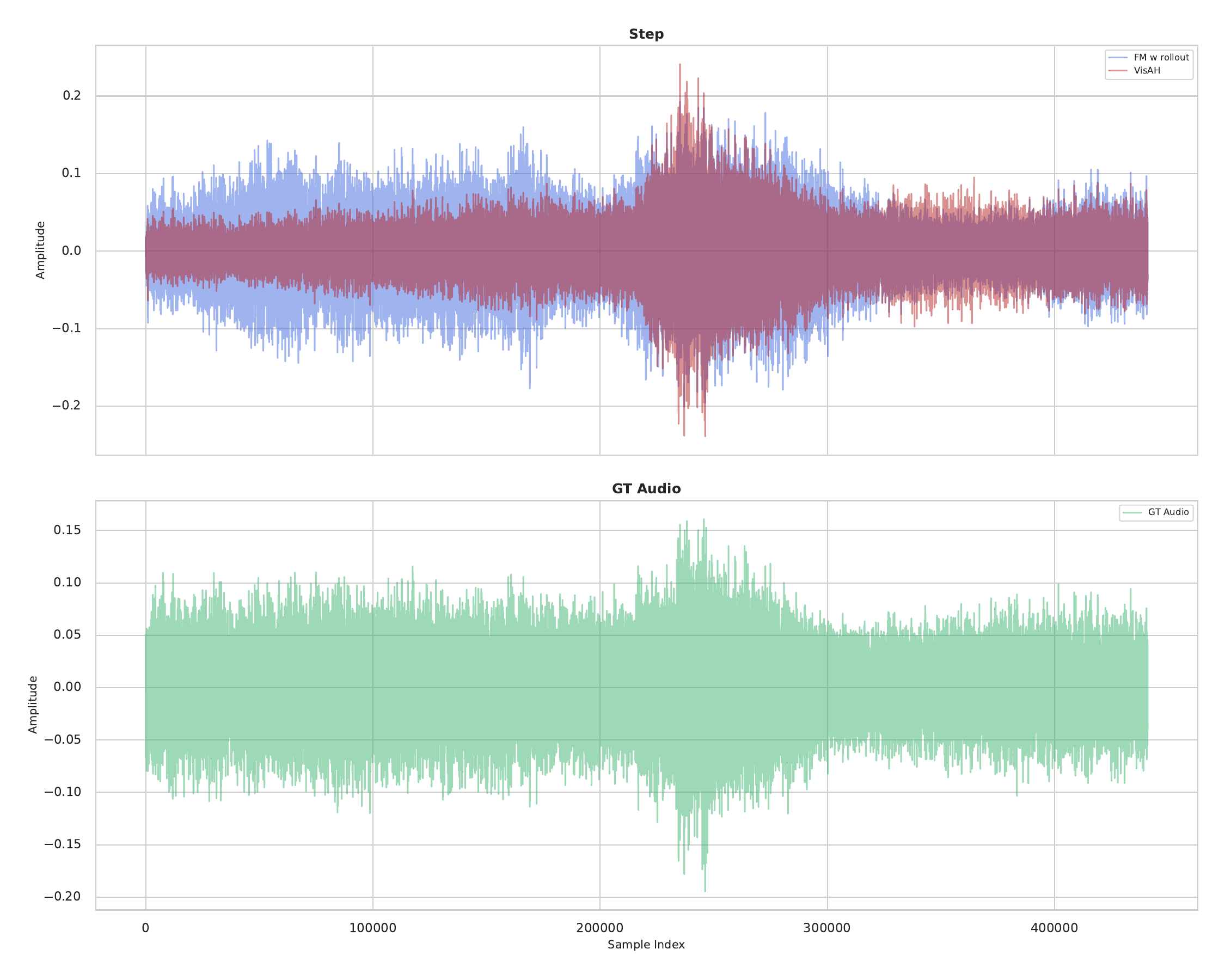}
        \label{fig:visah_sub4}
    \end{subfigure}

    \vspace{0.6em}

    % Row 3
    \begin{subfigure}{0.48\textwidth}
        \centering
        \includegraphics[width=\linewidth]{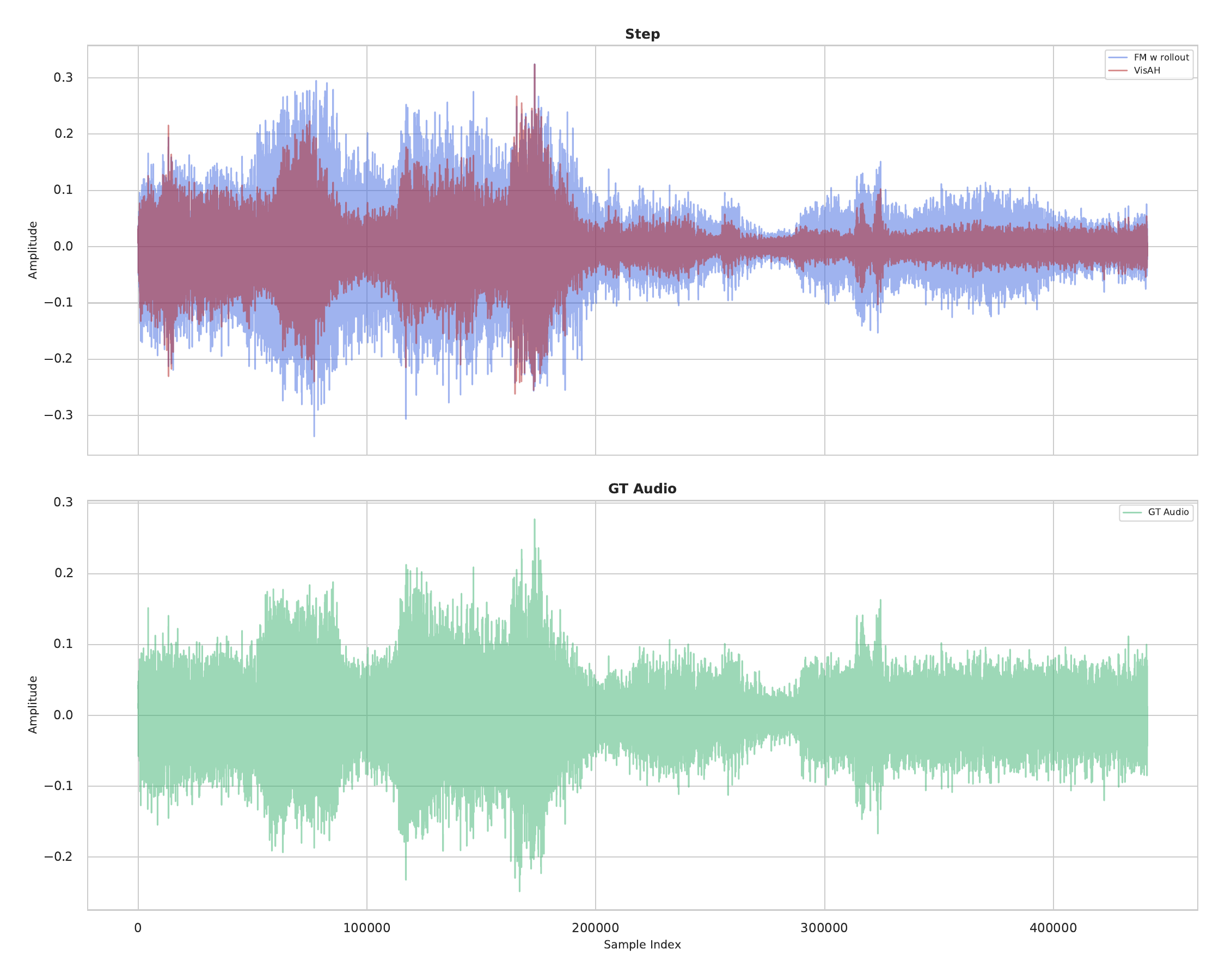}
        %\caption{Caption 5}
        \label{fig:visah_sub5}
    \end{subfigure}
    \hfill
    \begin{subfigure}{0.48\textwidth}
        \centering
        \includegraphics[width=\linewidth]{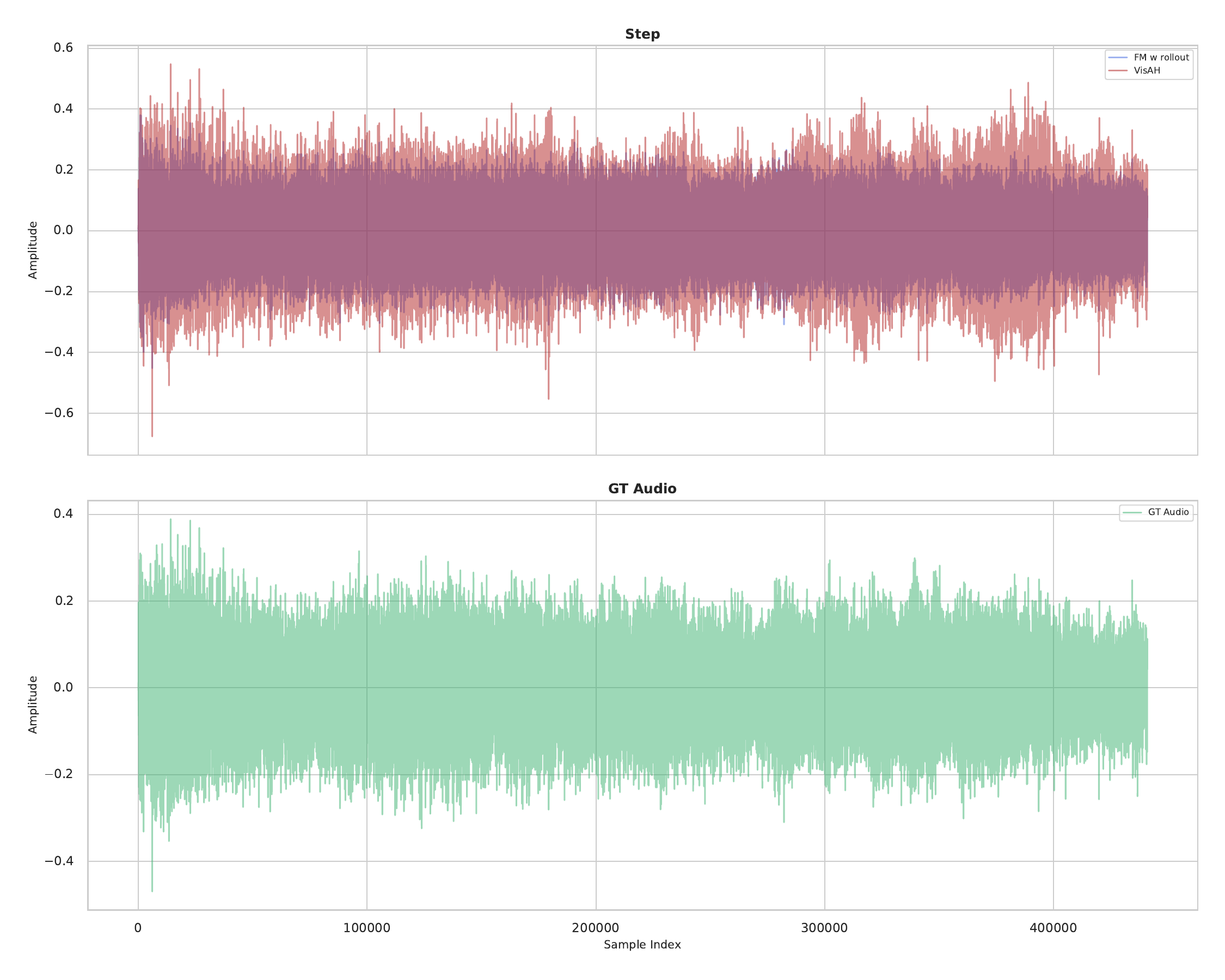}
        \label{fig:visah_sub6}
    \end{subfigure}

    \caption{Qualitative comparison between VisAH-FM and the discriminative VisAH.}
    \label{fig:fullVisAH}
\end{figure*}

Figure \ref{fig:fullRoll} shows further qualitative samples that highlight the difference of behavior between the flow matching models trained with and without rollout loss. The rollout loss allows more consistent predictions across steps, resulting in more highlighted sources.
Figure \ref{fig:fullVisAH} shows qualitative comparison between VisAH and our model VisAH-FM. Our model find more accurately the source to enhance and emphasize it more. 
Finally, audio samples can be found in attached zip file.

.
\end{document}